\documentclass[
  journal=largetwo,
  manuscript=article-type,
  year=2024,
  volume=39,
]{cup-journal}

\newcount\ppnum
\newcommand\ppnumber[1]{%
        \ppnum=#1\relax
        \ifnum\ppnum<0
                $-$%
                \ppnum=-\ppnum
        \fi
        \let\pptemp\empty
        \loop\ifnum\ppnum>999
                \count255=\ppnum
                \divide\ppnum by1000
                \count255=\numexpr \count255 - 1000*\ppnum \relax
                \edef\pptemp{,\ifnum\count255<100 0\ifnum\count255<10 0\fi\fi
                    \the\count255 \pptemp}%
        \repeat
        \the\ppnum
        \pptemp
}

\usepackage{biblatex}
\usepackage{amsmath, hhline}
\usepackage[nopatch]{microtype}
\usepackage{booktabs}
\graphicspath{{Figures/}}
\usepackage{caption,subcaption}
\usepackage[]{hyperref}
\newcommand{\teff}{$T_{\rm eff}$}
\newcommand{\logg}{$\log g$}
\newcommand{\vsini}{$v \sin i$}

\newcommand{\ds}{$\delta$ Scuti}
\hypersetup{colorlinks = true, linkcolor = blue, anchorcolor = red, citecolor = blue, filecolor = red, urlcolor = red, hypertexnames=true}

\def\AE#1{\textcolor{Green}{#1}}

\def\VB#1{\textcolor{cyan}{#1}}

\title{Absolute Parameters of Young Stars: NO Puppis}

\author{Ahmet Erdem}
\affiliation{Astrophysics Research Center \& Ulup{\i}nar Observatory, \c{C}anakkale Onsekiz Mart University, TR-17100, \c{C}anakkale, T\"{u}rkiye}
\alsoaffiliation{Department of Physics, Faculty of Science, \c{C}anakkale Onsekiz Mart University, Terzio\u{g}lu Kamp\"{u}s\"{u}, TR-17100, \c{C}anakkale, T\"{u}rkiye}

\author{Volkan Bak{\i}\c{s}}
\affiliation{Department of Space Sciences and Technologies, Faculty of Sciences, Akdeniz University, 07058 Antalya, T\"{u}rkiye}
\email[Volkan Bak{\i}\c{s}]{volkanbakis@akdeniz.edu.tr}

\author{John Southworth}
\affiliation{Astrophysics Group, Keele University, Staffordshire, ST5 5BG, UK}

\author{Michael D. Rhodes}
\affiliation{Brigham Young University, Provo, Utah 84602, USA}

\author{Filiz Kahraman Ali\c{c}avu\c{s}}
\affiliation{Astrophysics Research Center \& Ulup{\i}nar Observatory, \c{C}anakkale Onsekiz Mart University, TR-17100, \c{C}anakkale, T\"{u}rkiye}
\alsoaffiliation{Department of Physics, Faculty of Science, \c{C}anakkale Onsekiz Mart University, Terzio\u{g}lu Kamp\"{u}s\"{u}, TR-17100, \c{C}anakkale, T\"{u}rkiye}

\author{Edwin Budding}
\affiliation{Carter Observatory, 40 Salamanca Road, Kelburn, Wellington 6012, New Zealand}
\alsoaffiliation{School of Chemical \& Physical Sciences, Victoria University of Wellington, PO Box 600, Wellington 6140, NZ}

\author{Mark Blackford}
\affiliation{Variable Stars South, Congarinni Observatory, Congarinni, NSW, 2447, Australia}

\author{Timothy Banks}
\affiliation{Department of Physical Science \& Engineering, Harper College, 1200 W Algonquin Rd, Palatine, IL 60067, USA}
\alsoaffiliation{Nielsen, 675 6th Ave, New York, NY 10011, USA}

\author{Murray Alexander}
\affiliation{Physics Department, University of Winnipeg, 515 Portage Avenue, Winnipeg R3B 2E9, Canada}

\keywords{keyword entry 1, keyword entry 2, keyword entry 3} 

\begin{document}

\begin{abstract}
The southern early-type, young, eccentric-orbit eclipsing binary NO Puppis forms the A component of the multiple star Gaia DR3 552\-8147999779517568. The B component is an astrometric binary now at a separation of about 8.1 arcsec.  There may be other fainter stars in this interesting but complex stellar system. We have combined several lines of evidence, including TESS data from 4 sectors, new ground-based BVR photometry, HARPS (ESO) and HERCULES (UCMJO) high-resolution spectra and astrometry of  NO Pup.  We derive a revised set of absolute parameters with increased precision. Alternative optimal curve-fitting programs were used in the analysis, allowing a wider view of modelling and parameter uncertainties. The main parameters are as follows: $M_{Aa} = 3.58 \pm 0.11$, $M_{Ab} = 1.68 \pm 0.09$ (M$_\odot$); $R_{Aa} = 2.17 \pm 0.03$, $R_{Ab} = 1.51 \pm 0.06$ (R$_\odot$), and $T_{\rm e Aa} = 13300 \pm 500$,  $T_{\rm e Ab} = 7400 \pm 500$ (K). We estimate approximate masses of the wide companions, Ba and Bb, as $M_{Ba} = 2.0$ and $M_{Bb} = 1.8$ (M$_\odot$). 
The close binary's orbital separation is $a= 8.51 \pm 0.05$  (R$_\odot$); its age is approximately $20$ Myr and distance $172 \pm 1$ pc.  The close binary's secondary (Ab) appears to be the source of low amplitude  $ {\delta}$ Scuti-type oscillations, although the form of these oscillations is irregular and unrepetitive.    Analysis of the $ \lambda$ 6678 He I profile of the primary show synchronism of the mean bodily and orbital rotations.  The retention of significant orbital eccentricity, in view of the closeness of the A-system components, is unexpected and poses challenges for the explanation that we discuss.
\end{abstract}


\section{Introduction} 
\label{sec:intro}

This paper forms part of a programme addressing relatively neglected  close binary systems in the Southern Hemisphere \citep{Idaczyk_2013}. A recent example was that of \citet{Erdem_2022} dealing with the system V410 Pup, which has some similarities in its physical properties to NO Pup. Close binary systems provide a recognised source of fundamental data on stellar parameters, notably their masses and radii. The advent of space based facilities, such as the Transiting Exoplanet Survey Satellite (TESS: \citeauthor{Ricker_2014}, \citeyear{Ricker_2014} and \citeyear{Ricker_2015}) with its wealth of high-precision photometry, together with available high-resolution spectrometry, have transformed our knowledge of stellar astrophysics.

\begin{figure}[!t]
    \begin{center}
        \includegraphics[scale=0.35]{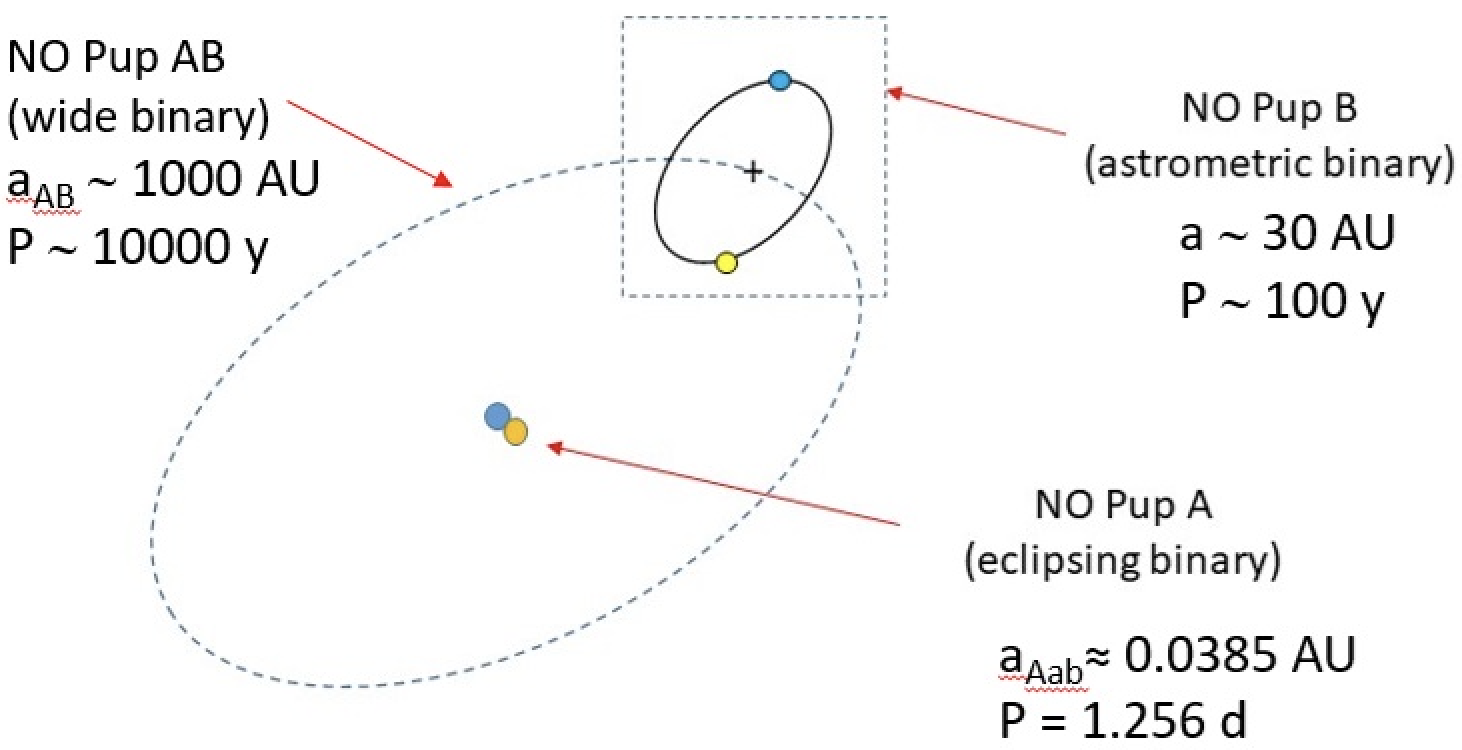} 
    \end{center}
\caption{Schematic of the main four stars of the NO Pup system.} 
\label{fig_system}
\end{figure}

This knowledge becomes interestingly augmented in the case of eclipsing binaries such as NO Pup (HD 71487, HIP 41361, CoD --38$^{\circ}$4462, HR 3327), whose orbital motion includes a steady additional rotation of its main elliptical form. This is associated with the departure from simple Keplerian motion mainly due to the perturbation from the sphericity of the component stars. Data on the difference between observed and calculated times of minimum light from these eclipsing binaries,  `O -- C's, allow tests  of the physics underlying this apsidal motion \citep{Wolf_2008}.  Substantiation of this point can be found in the reviews of \citet{Giminez_1992}, \citet{Claret_1993}, \citet{Tohline_2002}, \citet{Horch_2013}, and others. The short-period ($\sim$1.2569 d) of NO Pup's orbit makes for a relatively fast apsidal motion.  In  fact, that such a short period system should have retained a noticeable eccentricity seems  {\em prima facie} surprising (cf.\ the case of $\zeta$~TrA; \cite{Skuljan_2004}).  As well as the classical two-body discussion of apsidal motion \citep{Sterne_1939}, the case of NO Pup may present effects associated with the close binary's wide orbit companions.

The early-type close pair forms the `A' component of the multiple star WDS J08263-3904\footnote{Gaia DR3 552\-8147999779517568}  (Fig~\ref{fig_system}). The `B' component, discovered  by John Herschel in 1835, is now at a separation of about 8.1 arcsec and position angle 124 deg. The AB double star, at a distance of 172 pc, has apparently closed in by a few arcsecs over the last $\sim$ 200 yr, although without a significant change of position angle.  The B component is itself an astrometric binary (Bab), with a relatively short period of 103 yr.  This pair has a combined V magnitude of about 7, and the source would have been included in conventional photometry of NO Pup A. A third component (C)  apparently has optical closeness only \citep{Veramendi_2014}, while companion D was identified by \citet{Tokovinin_1999}, with a separation of 5.4 arcsec and position angle 265$^{\circ}$. The whole group lies to the south of a small asterism, about half a degree south of the Galactic plane.

The close pair's combined brightness is given as  V $\approx 6.49$ (SIMBAD), with $\rm B-V \approx -0.06$, which is slightly bluer than the reported combination of Main Sequence spectral types \AE{B8V}+A7V would suggest (for the Aa and Ab stars). The visible companion Bab is reported with V mag 7.04 and $\rm B - V = 0.07$ (SIMBAD: \cite{Wenger_2000}). 

NO Pup (Aab) was discovered to be variable by \citet{Gronbech_1976} during $uvby$ photometry of bright southern stars in 1972 \citep{Jorgensen_1972}. He derived the first light elements,  revealing the relatively short orbital period.  A diaphragm of 30 arcsec was used in the photometric measurements of \citet{Gronbech_1976}, so the light contribution of the visible companion was always included in the flux measurements.  This four-colour photometry  was analysed by \citet{Gimenez_1986}, who gave the following linear ephemerides for the primary minimum:
\begin{align}
    {\rm Min\ I} = \ & {\rm HJD} \, 2441752.6576 \pm 0.0012  \ + \nonumber \\
                     & ( 1.2569 \pm 0.0027 ) E
\end{align}
and for the secondary:
\begin{align}
    {\rm Min\ II} = \ &{\rm HJD} \, 2441753.3780 \pm 0.0004 \ +  \nonumber \\
                    &( 1.2569 \pm 0.0004 ) E
\end{align}
\citet{Gimenez_1986} improved our knowledge of the orbit and its behaviour, announcing the short period of apsidal motion $U = 37.2 \pm 0.2$ yr. 

\vfill\eject  
\section{Photometry}
\label{sec:photometry}

\subsection{WinFitter fits to TESS data}
\label{sec:wf_tess_fits}

\begin{figure}[!t]
        \begin{center}
            \includegraphics[width=1.0\columnwidth]{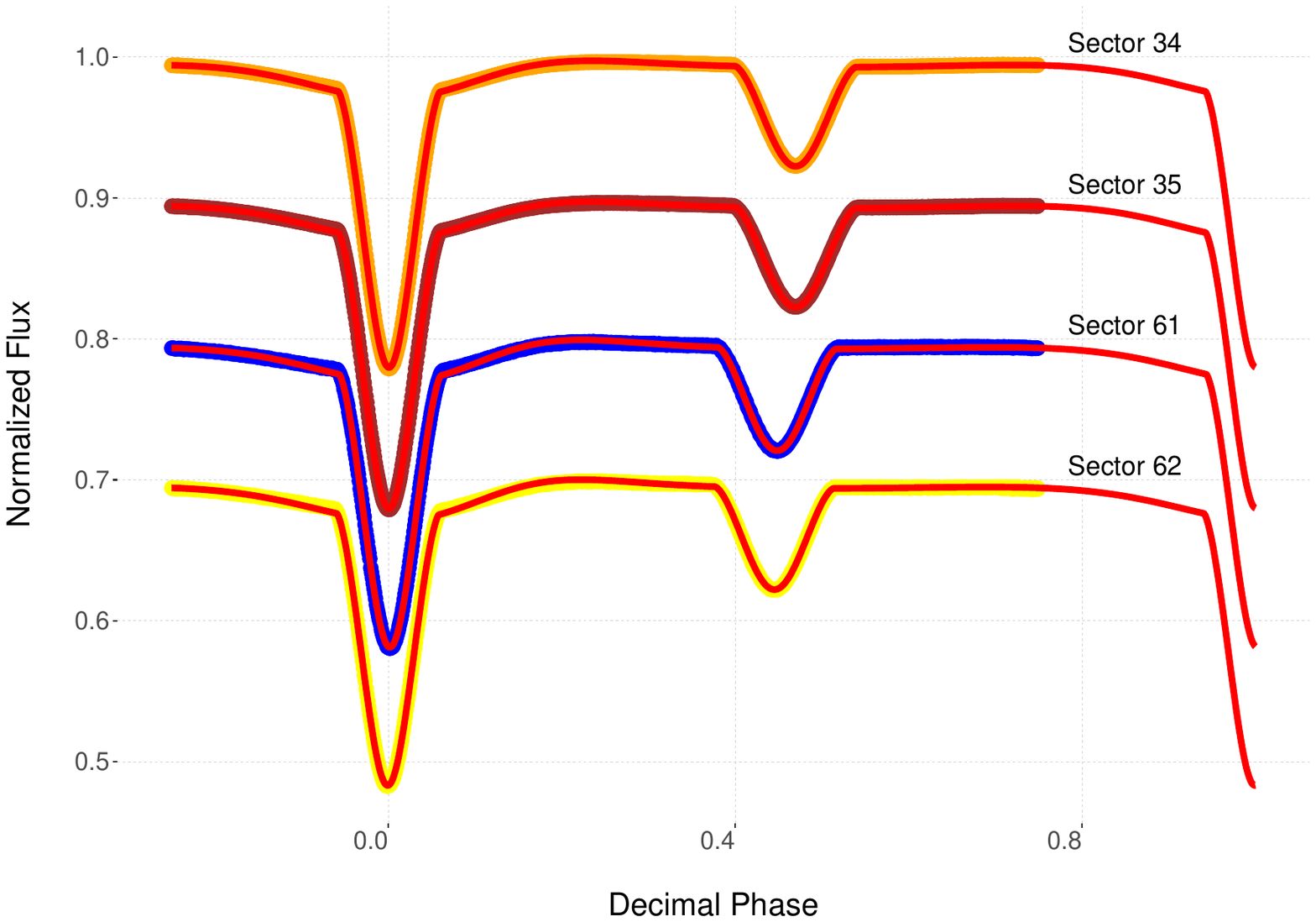}
        \end{center}
\caption{Binned TESS data are plotted for sectors 34, 35, 61, and 62 with optimal {\sc WinFitter} models. Fluxes for Sector 35 are offset by $-0.2$ from their actual values, Sector 61 by a further $-0.2$, and Sector 62 by an additional $-0.2$.  The model fluxes are presented as red continuous curves. The TESS data for each sector have been folded by the orbital period and then binned to 3600 points.  Model parameters are given in Table~\protect\ref{tab:winfitter_tess_models}.}
\label{fig:tess_winfitter_fits}
\end{figure}

\begin{table*}
    \caption{Parameter values for {\sc WF} models to TESS Sectors 34, 35, 61, and 62, folding each sector's data by the ephemeris given by \citet{Veramendi_2014}. The mass ratio $q$ was adopted as 0.47 \citep{Veramendi_2014}. The linear limb darkening coefficient for the primary star was set at 0.29, and  for the secondary  0.39.
{
   The latter parameters depend on the assigned effective temperatures and wavelength.
   These were set as $T_1 = 12000$ K; $T_2 = 7700$ K; $\lambda_{\rm eff} = 0.835 \mu$m.  
   The fractional luminosities $L_i$
   are the mean relative fluxes from each star,
   normalized so that their sum is unity. The radii $r_i$ are mean radii of the two stars in the close binary system divided by the semi-major axis of the relative orbit. Angles are given in degrees.
   $M_0$ is the mean anomaly at phase zero.
   The phase bin size is then close to 0.4 deg.}  
   See Figure~\protect\ref{fig:tess_winfitter_fits} for plots of the model fits to the four data sets.}
    \centering
    \begin{tabular}{l|l|l|l|l}
    \hline
    Parameter                               & Sector 34                 &  Sector 35                    & Sector 61               & Sector 62                \\
    \hline
    Primary luminosity $L_1$                & $0.576 \pm 0.015$           & $0.608 \pm 0.015 $             & $0.602 \pm 0.014$         & $0.595 \pm 0.012$          \\
    Secondary luminosity $L_2$              & $0.123 \pm 0.010$         & $0.114 \pm 0.008 $            & $0.140 \pm 0.008$       & $0.134 \pm 0.007$        \\
    Tertiary luminosity $L_3$               & 0.301 $\pm 0.019$           & $0.278 \pm 0.018 $         & $0.258 \pm 0.020$         & $0.271 \pm 0.016$          \\  
    Primary radius $r_1$                    & $0.254 \pm 0.001$         & $0.255 \pm 0.001 $            & $0.256 \pm 0.002$       & $0.254 \pm 0.001$        \\
    Secondary radius $r_2 $                 & $0.181 \pm 0.003$         & $0.177 \pm 0.002 $            & $0.184 \pm 0.003$       & $0.181 \pm 0.002$        \\ 
    Inclination $i$                         & $78.3 \pm 0.6  $          & $78.6 \pm 0.7 $               & $79.9 \pm 0.8$          & $78.4\pm 0.6$           \\
    Eccentricity $e$                        & $0.131 \pm 0.012$         & $0.132\pm 0.013$              & $0.141 \pm 0.011$       & $0.139 \pm 0.013$        \\
    Mean anomaly $M_0$                      & $343.7 \pm 6.9$           & $ 343.8\pm 6.6$                & $330.6 \pm 6.7 $        & $343.7 \pm 7.0$          \\
    Exposure time (seconds)                 & 120                       & 120                           & 120                     & 120                      \\
    \hline
    \end{tabular}
    \label{tab:winfitter_tess_models}
\end{table*}

{
We downloaded the TESS data in sectors 34, 35, 61 and 62 with short cadence (120-s sampling) from the Mikulski Archive for Space Telescopes (MAST).  Our adopted procedure for
extraction of TESS data has been spelled out by Blackford (2025).
While noting the Pre-search Data Conditioning Simple Aperture Photometry (PDCSAP) \citep{Jenkins_etal_2016},
we concentrated LC analysis on the Simple Aperture Photometry (SAP) fluxes, since the PDCSAP detrending produces artificial side-effects associated with the search for planetary transits.}

{
LC modelling for close binary stars often refers to the numerical integration procedure of \citet{Wilson_1971} (WD), which represents the distorted component surfaces as equipotentials, according to the classical point-mass formulation attributed to \citet{Roche_1873}, recalled in Ch.\ 3 of \citet{Kopal_1959}. Both the WD and WiNFitter (WF) methods converge to the same approximation for the surface perturbation when the internal structural constants $k_j$ are neglected, implying disregard of the effects of tides on tides.
The relevant formula,  Eqn 1-11 in Ch 2, or Eqn 2-6 in Ch. 3 of \citet{Kopal_1959}, is:
\begin{equation}
    \frac{ \Delta^{\prime} r }{r_0}  = q \sum^{4}_{j = 2} r_0^{j+1} (1 + 2k_j) P_j(\lambda) + n r_o^3(1 - \nu^2)  \,\,\, ,
    \label{eq:fosp}
\end{equation}
where $r$ is the local stellar radius expressed as a fraction of the orbital separation of the components with mean value $r_0$.
{$ P_j(\lambda)$ is the Legendre polynomials and $n = (1+q)/2$, where $q$ is the mass ratio. $\lambda$ and $\nu$ are the direction cosines for an arbitrary point on the star's surface with respect to the line of centres ($\lambda$) and the spin axis ($\nu$). } 
In any case, the difference between the stellar distortions in WD and WF become small at greater relative separations of the two stars ($r_0 \rightarrow 0.$).
{The direction cosine of the angle between the radius vector $\hat{r}$ and the line of centres is here $\lambda$, and $\nu$ is the direction cosine of the angle between $\hat{r}$ and the rotation (`spin') axis. } 
The coefficients $k_j$ (in WF)  can be taken from suitable stellar models, e.g.\ \citet{Inlek_2017}. They are set to zero in WD.

Sector 34 gathered observations of the photometric flux during the period 2021 Jan 14 to 2021 Feb 08; Sector 35 2021 Feb 09 to 2021 Mar 06; Sector 61 2023 Jan 18 to 2023 Feb 12; and Sector 62 2023 Feb 12 to 2023 Mar 10.\footnote{https://heasarc.gsfc.nasa.gov/docs/tess/sector.html} \citet{Rhodes_2023} provides background and a user manual for {\sc WinFitter}, \citet{Banks_1990}  discussed its optimization methods that build on chapter 11 of \citet{Bevington_1969}.   {\sc WinFitter} numerically inverts the Hessian of the $\chi^2$ variate in the vicinity of its minimum to derive estimates for resulting parameter uncertainties, that include effects of inter-correlations between these parameters. 

Later, we binned the \ppnumber{17385} individual observations of Sector 34 by phase to produce \ppnumber{1021} representative points. Similarly, \ppnumber{14156} observations in Sector 35 were binned to \ppnumber{1010} points, \ppnumber{17771} to \ppnumber{1045} in Sector 61, and \ppnumber{18025} to \ppnumber{1045} in Sector 62.  Such binning tends to remove photometric structure with periods not synchronised with that of the orbit (see Section~\ref{sec:wd_fits}). However, certain systematic, but unmodelled effects remain in the data, working against the idea of one clear and unequivocal LC `solution'. Parameter estimates from the best fitting model are given in Table~\ref{tab:winfitter_tess_models},  while Figure~\ref{fig:tess_winfitter_fits} displays the model fits to each sector's data.

{The main optimal parameters derived from the WF analysis of the TESS light curves are as follows (the angular parameters $i$, $M_0$ and $\omega$ in degrees) : 
$L_1 = 0.595\pm 0.013$,  $L_2 = 0.128\pm 0.008$,  $L_3 = 0.277\pm 0.019$, $r_1 = 0.255\pm 0.001$, $r_2 = 0.181\pm 0.004$,  $i = 78.8 \pm 0.4$, $e = 0.136\pm 0.012$,  $M_0 = 340.5\pm 6.6$, $\omega = 119.8 \pm8.5$. }



\subsection{WinFitter fits to BVR Photometry}
\label{sec:wf_bvr_fits}

Multicolour photometry of NO Puppis was carried out over 6 nights in January and February 2019 from the Congarinni Observatory, NSW, Australia (152° 52’ E, 30° 44’ S, Alt.\ 20 m). Images were captured with an ATIK$^{\rm TM}$ One 6.0 CCD camera equipped with Johnson–Cousins BVR filters attached to an 80 mm f6 refractor stopped down to 50 mm aperture. MaxIm DL$^{\rm TM}$ software was used for image handling, calibration, and aperture photometry. 

HD 71932 was the main comparison star. Its magnitude and colours were measured as V = 8.274, $\rm B - V$ = 0.486, and $\rm V - R$ = 0.283. The magnitudes and colours of NO Pup just before and after primary eclipse were then determined as V = 6.086(6), B -- V = { --0.019(10), V -- R = --0.002(9)}.  These measures, coupled with the determinations of the relative fluxes of the components from the LC analysis,   allow the magnitudes and colours of the components to be derived, and thence  the surface temperatures checked.

Table~\ref{tab:winfitter_bvr_models} 
presents the best fit parameter estimates for the three light curves, while Figure~\ref{fig:bvr_winfitter_fits} 
shows the model LC fits to the observations. Table~\ref{tab:mark_c_separate} 
presents the magnitudes of the components in the standard BVR system. From this we see that the primary star  (NO Pup Aa) has B -- V of $-0.11$, in agreement with the assigned  B8V spectral type (Section \ref{sec:intro}). Ab, with B -- V = 0.30, corresponds to an F0 main sequence star ( Table 9.2 in \cite{Budding_2022}). We could expect NO Pup B to be characterised by a mid-A dwarf spectral type.  The WDS catalogue gives a slightly brighter combination V magnitude for NO Pup B as 7.23. The separate magnitudes given in the same catalogue would correspond to  A5V and A6V spectral types, with a total mass of about 3.8 $M_{\odot}$.

\begin{table}[!t]
    \caption{BVR magnitudes of NO Pup Aab and B.}
    \centering
    \begin{tabular}{c|r|r|r|r}
    \hline
          & \multicolumn{4}{c}{NO Pup components} \\
          Filter & mag Aa & mag Ab & mag B    & Err. (mag Aa) \\
    \hline
    $B$ & 6.479  & 9.26 & 7.51       & 0.019 \\  
    $V$ & 6.588  & 8.96 & 7.40       & 0.015 \\
    $R$ & 6.593  & 8.91 & 7.40       & 0.013 \\ 
    \hline
    \end{tabular}
    \label{tab:mark_c_separate}
\end{table}

\begin{figure}[!t]
        \begin{center}
            \includegraphics[width=1.0\columnwidth]{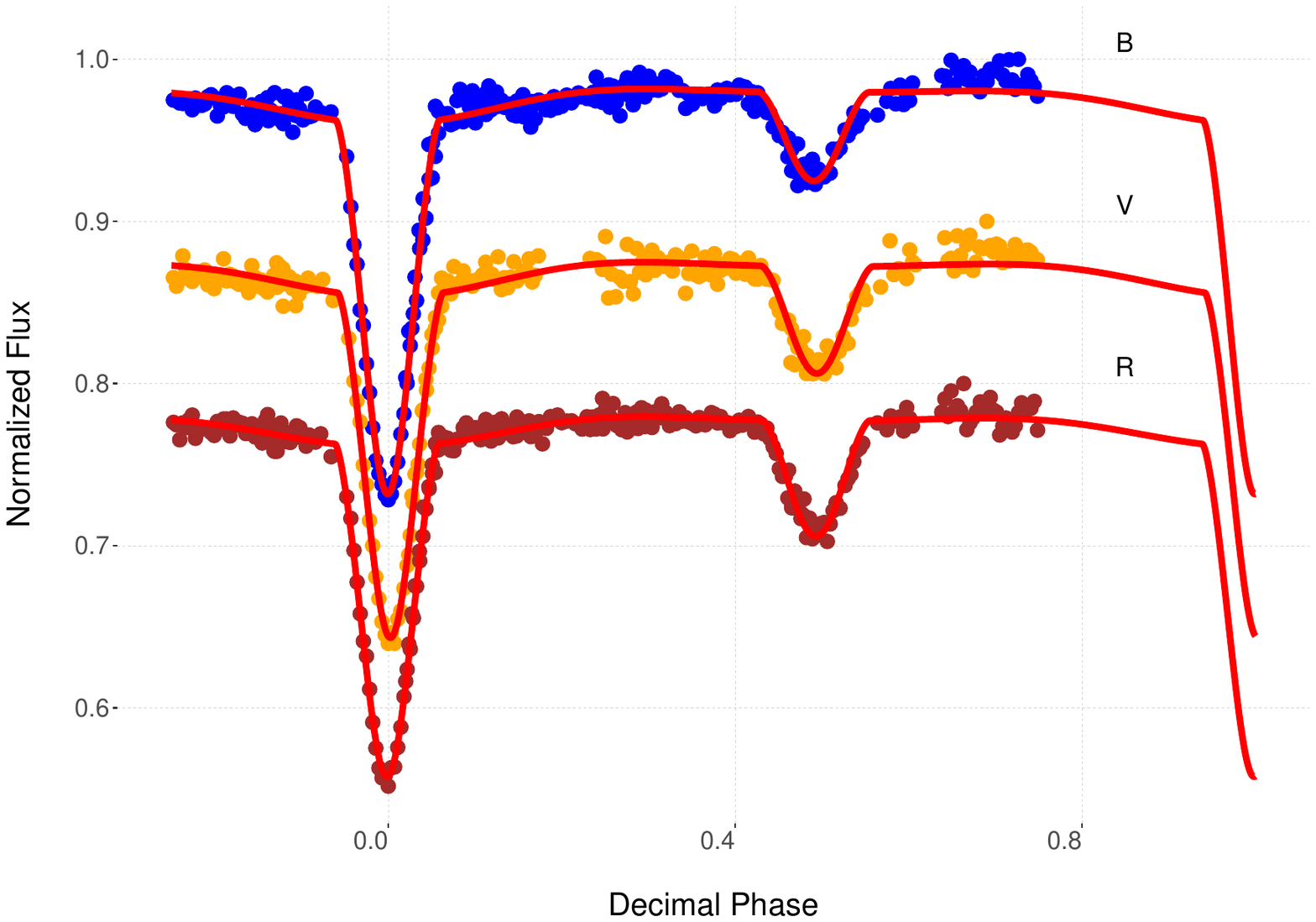}
        \end{center}
\caption{{\sc WinFitter} model lightcurves for the ground-based {\em BVR} photometry. The $V$ and $R$ light curves are offset by $-0.1$ and {-0.2} respectively in normalised flux for display purposes. Optimal parameter values are listed in Table~\protect\ref{tab:winfitter_bvr_models}}
\label{fig:bvr_winfitter_fits}
\end{figure}

\begin{table*}
    \caption{Parameter values for {\sc WinFitter} models to the {\em BVR} photometry. 
    {The parameter symbols carry the same meaning as in Table~\ref{tab:winfitter_tess_models}.}
    See Figure~\protect\ref{fig:bvr_winfitter_fits} for plots of the model fits to the three data sets. Angles are in degrees.  { The eccentricity {\color{blue}  ($e = 0.127$)}, adopted after checking 
    the results of numerous optimisation estimates,
    has been used in these fittings (see Section~\ref{sec:absolute_parameters}).} }
    \centering
    \begin{tabular}{l|l|l|l}
    \hline
    Parameter                               & B                         &  V                            & R                   \\
    \hline
    Primary Luminosity $L_1$                & $0.684 \pm 0.079$         & $0.630 \pm 0.020 $            & $0.628     \pm 0.089$   \\
    Secondary Luminosity $L_2$              & $0.053 \pm 0.008$         & $0.071 \pm 0.002 $            & $0.074 \pm 0.009$   \\
    Tertiary Luminosity $L_3$               & $0.264 \pm 0.093$         & $0.299 \pm 0.027 $            & $0.298 \pm 0.100$   \\  
    Primary radius $r_1$                    & $0.261 \pm 0.009$         & $0.247 \pm 0.003 $            & $0.244 \pm 0.010$   \\
    Secondary radius $r_2$                  & $ 0.183 \pm 0.027$        & $0.163 \pm 0.004$             & $0.151 \pm 0.028$   \\
    Inclination $i$                         & $81.0 \pm 1.0$            & $80.8 \pm 0.4 $               & $81.4 \pm 0.9$      \\
    Mean Anomaly $M_0$                      & $355.8 \pm 0.8$            & $356.2 \pm 0.7 $              & $357.7 \pm 0.6$       \\
    Exposure time (seconds)                 & 30                        & 20                            & 30                 \\
    \hline
    \end{tabular}
    \label{tab:winfitter_bvr_models}
\end{table*}


\subsection{WD+MC fits to TESS data}
\label{sec:wd_fits}

We separately downloaded TESS data in sectors 34, 35, 61 and 62 in short cadence (120-s sampling rate) from the MAST\footnote{https://mast.stsci.edu/portal/Mashup/Clients/Mast/Portal.html}, as recalled in Subsection \ref{sec:wf_tess_fits}. 
We preferred to work with the SAP fluxes, as mentioned above.
Data with a quality flag of zero were selected and used approximately \ppnumber{17000}, \ppnumber{14000}, \ppnumber{16500} and \ppnumber{17000} points to define the LCs for sectors 34, 35, 61 and 62, respectively. 

In the light curves of NO Pup, the light level of maximum I remain higher than that of maximum II. 
The maximum I light level was then selected for reference in all sectors of the TESS data and additional detrending was applied with a low-order polynomial fit in order to normalise the LCs. 
As an example, the maximum light levels in two consecutive orbital light curves selected from Sector 35 are shown in Figure \ref{fig_example_lc}. Apart from the asymmetry in the maximum light levels, NO Pup shows low amplitude pulsations. We have performed a frequency analysis of these oscillations that will be discussed in Section~\ref{sec:pulsations}.

\begin{figure}[!t]
\centering
	\includegraphics[scale=0.27]{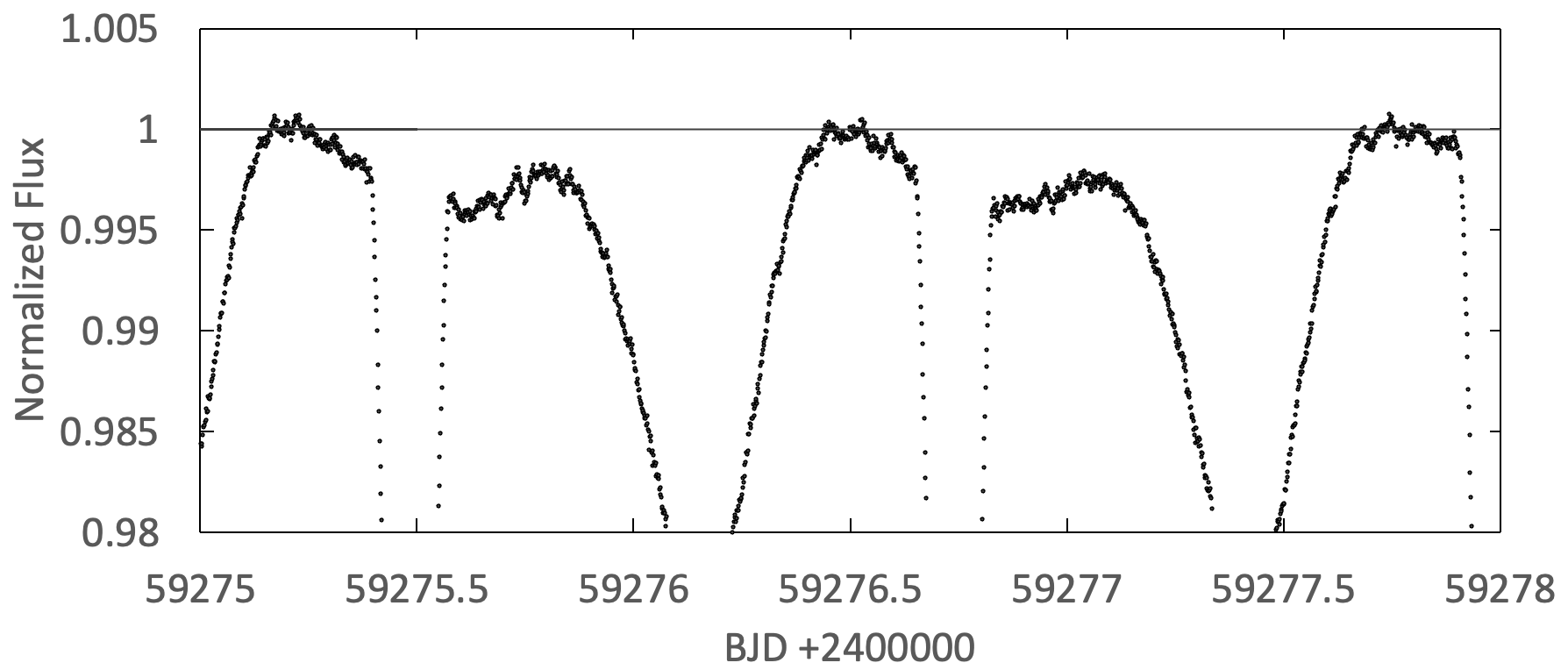}
\caption{Maximum light levels in the light curve of NO Pup along two consecutive orbits from TESS Sector 35 data. Asymmetry between the maximum light levels is evident and it is also seen that NO Pup shows pulsations with very low amplitude.} 
\label{fig_example_lc}
\end{figure}

In the SAP data  of all sectors the CROWDSAP parameter, i.e.\ the ratio of target to total flux in the photometric aperture, is given as 0.62. This indicates that approximately 38\% of the observed flux in the SAP aperture does not come from the target star. {This is most probably due to other background starlight collected in the rather large pixels of the TESS detector. The relative contribution of third light ($l_3$) was therefore taken into account in the LC analysis of NO Pup A (see Table~\ref{table_wd}).}

We used the numerical integration method of \citet{Wilson_1971} (WD) which models the light curve of a given binary star by taking into account proximity effects, regarding the surfaces of the components as Roche equipotentials. The original program, {\sc wd}, has been combined with a Monte Carlo (MC) optimisation procedure, as discussed in \citet{Zola_etal_2004}. Representative values and uncertainties of the adjusted parameters were derived in this way.

{\sc wd} requires a preliminary value of the primary's effective temperature, which was deduced as follows:
\citet{Veramendi_2014} assigned the spectral type of the system as B5V + B9.5V from their spectral analysis;. from that, they adopted that $T_{1}$ = 13000 K. In our later Section \ref{sect:atm} we set $T_{1}$ = 13500 and 13700 K (Table \ref{tab:atmospar})  by applying Kurucz atmospheric modelling to the disentangled spectrum of the primary. However, the colour index of B-V = --0.11 mag  from the photometric analysis in Section \ref{sec:wf_bvr_fits} yields the spectral type as B8V for the primary star. Furthermore, we estimated $T_{1}$ = 12000 K from the measured equivalent widths of He I 6678 lines in the UCMJO spectra (see Section \ref{subsection:ucmjo_spectra}). As a result, the values trialled for $T_{1}$ range from 12000 to 13700 K. 
 
{To determine which $T_{1}$ value is the most suitable, we followed the following procedure: TESS Sector 62 light curves were fitted with $T_1$ values between 11000 and 14000 K in steps of 500 K. We then calculated the photometric parallax from each LC solution.  As a result, we adopted the value $T_{1}$ = 13000 K in our LC analysis. This gives the photometric parallax value (5.747 mas, see Section \ref{sec:absolute_parameters}) closest to the Gaia DR3 parallax. Gaia DR3 gives the trigonometric parallax for NO Pup as 5.799 mas \citep{Gaia_2023}. If we add to this the 0.015 mas to account for the zero-point offset suggested by \citet{Lindegren_etal_2021}, the trigonometric parallax of NO Pup becomes 5.814 mas. The effective temperature of the secondary ($T_2$) was adjustable in the range of 5000 to 9000 K.}


The input range of the orbital inclination was set to $70^\circ < i < 90^\circ$, considering the {\sc WinFitter} LC fittings in Section \ref{sec:wf_tess_fits}.  For changes in $T_0$ and $P$, the input range for phase-shift ($\Delta \phi$; which allows the WD code to adjust for a zero-point error in the ephemeris used to compute the phases; the unit is the orbital period) was set to $-0.01 < \Delta \phi <0.01$. 

\begin{figure}[!t]
\centering
	\includegraphics[scale=0.48]{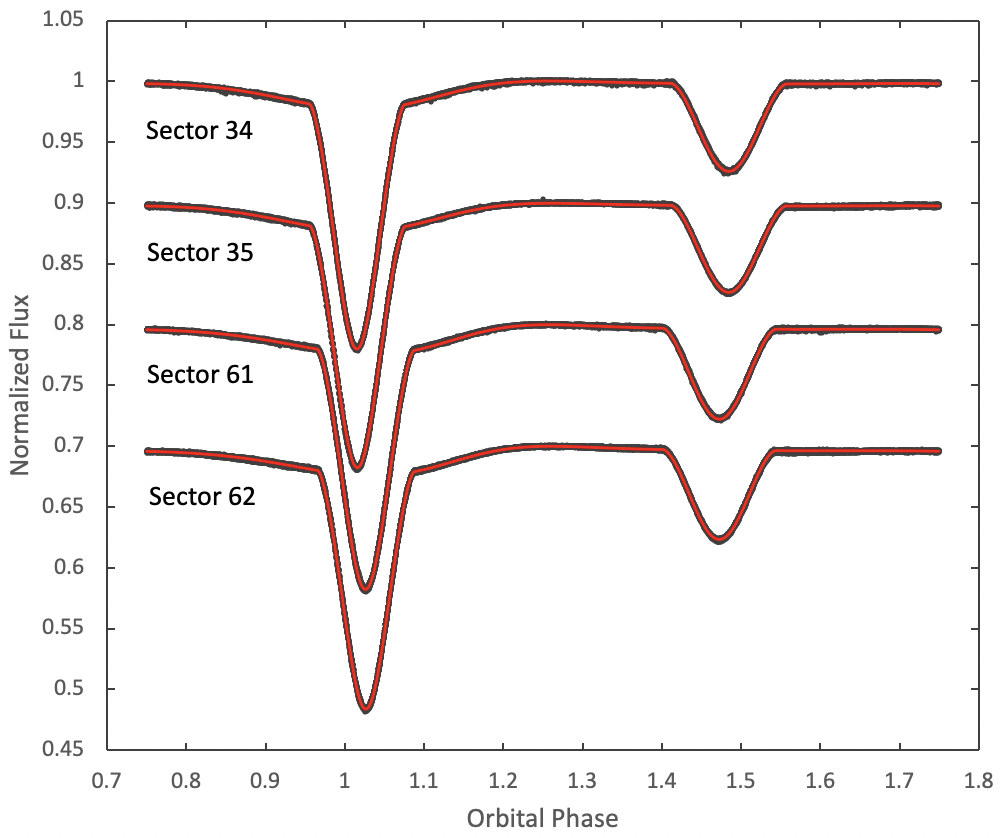}
	\includegraphics[scale=0.48]{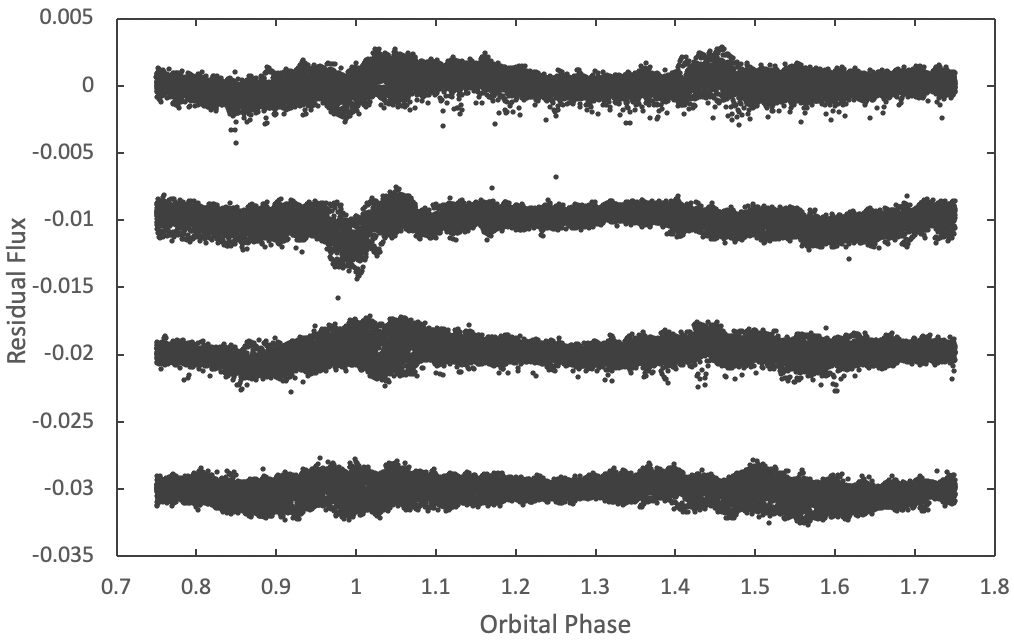} 
\caption{TESS light curves with the WD model fitting. Residuals to the LC model are plotted in the lower figure. The fluxes for sectors 35, 61 and 62 and their residuals are shifted downward to enhance
visibility.} 
\label{fig_lc_wd}
\end{figure}

\begin{figure}[!t]
\centering
	\includegraphics[scale=0.43]{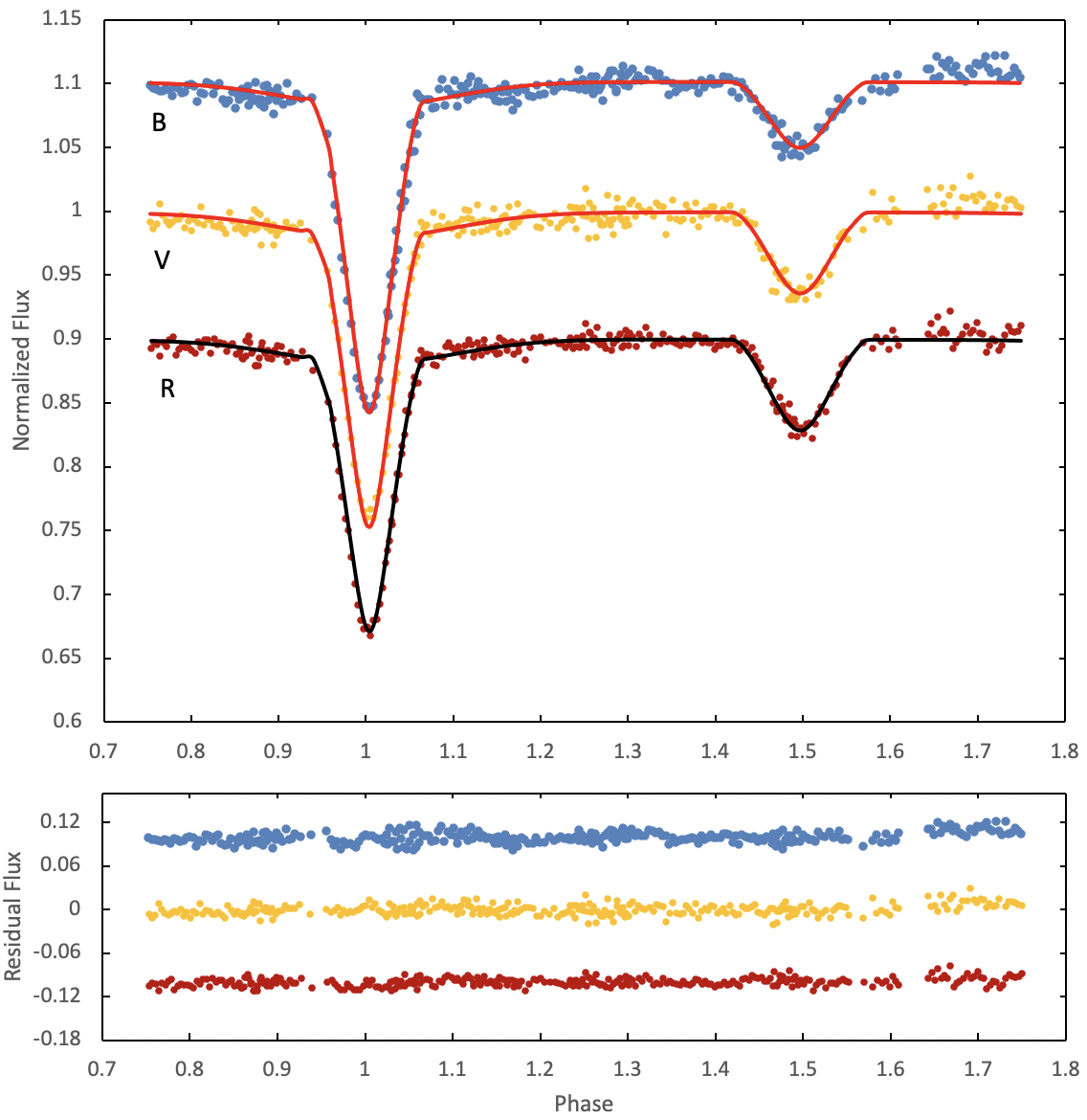}
\caption{BVR light curves with the {\sc wd+mc} model fitting. Residuals to the LC model are plotted in the lower figure.} 
\label{fig_bvr_wd}
\end{figure}

\begin{table*}[!tb]
    \caption{Final parameter values for WD+MC model to the BVR and TESS light curves. {$r$ (volume) is the radius of a sphere having the same volume as the tidally distorted star. $l_3$ is the third light contribution to the total light at phase 0.25.}}
    \centering
    \begin{tabular}{l|c|c|c|c|c|c}
    \hline
Parameter           & BVR (Model I)                         & BVR (Model II)                         & Sector 34             & Sector 35             & Sector 61             & Sector 62 \\
\hline
$e$                 & $0.130$ (fixed)                       & $0.130$ (fixed)                       & $0.1302 \pm0.0008$    & $0.1269 \pm0.0012$    & $0.1307 \pm0.0007$    & $0.1260 \pm0.0009$	   \\
$\omega$ (deg)      & $94 \pm1$                             & $94 \pm1$                             & $111 \pm1$            & $112 \pm1$            & $129 \pm2$            & $131.5 \pm0.9$	           \\
$i$ (deg)           & $83.54 \pm0.31$                       & $81.33$ (fixed)                       & $81.37 \pm0.20$       & $81.46 \pm0.22$       & $80.95 \pm0.21$       & $81.56 \pm0.18$	       \\
$T_1$ (K)           & $13000$ (fixed)                       & $13000$ (fixed)                       & $13000$ (fixed)       & $13000$ (fixed)       & $13000$ (fixed)       & $13000$ (fixed)	       \\
$T_2$ (K)           & $7516 \pm34$                          & $7454 \pm25$                          & $7306 \pm37$          & $7267 \pm42$          & $7380 \pm38$          & $7252 \pm39$	           \\
$q=M_{2}/M_{1}$	    & $0.473$ (fixed)                        & $0.473$ (fixed)                        & $0.473$ (fixed)        & $0.473$ (fixed)        & $0.473$ (fixed)        & $0.473$ (fixed)	        \\
$\Omega_1$          & $4.952 \pm0.043$                      & $4.390$ (fixed)                       & $4.403 \pm0.014$      & $4.406 \pm0.016$      & $4.375 \pm0.026$      & $4.387 \pm0.021$ 	       \\
$\Omega_2$          & $3.725 \pm0.043$                      & $4.336$ (fixed)                       & $4.321 \pm0.036$      & $4.328 \pm0.042$      & $4.334 \pm0.044$      & $4.378 \pm0.050$ 	       \\
$r_1$ (volume)        & $0.229 \pm0.010$                      & $0.264$ (fixed)                       & $0.263 \pm0.006$      & $0.262 \pm0.007$      & $0.265 \pm0.009$      & $0.263 \pm0.006$ 	       \\
$r_2$ (volume)        & $0.208 \pm0.008$                      & $0.163$ (fixed)                       & $0.164 \pm0.008$      & $0.163 \pm0.009$      & $0.163 \pm0.008$      & $0.160 \pm0.010$	       \\
$L_1/(L_1+L_2)$     & $0.86$, $0.82$, $0.79$ ($\pm0.12$)    & $0.93$, $0.91$, $0.89$ ($\pm0.09$)    & $0.882 \pm0.043$        & $0.884 \pm0.059$        & $0.882 \pm0.038$        & $0.889 \pm0.042$	       \\
$L_2/(L_1+L_2)$     & $0.14$, $0.18$, $0.21$ ($\pm0.05$)    & $0.07$, $0.09$, $0.11$ ($\pm0.01$)    & $0.118 \pm0.008$        & $0.116 \pm0.009$        & $0.118 \pm0.007$        & $0.111 \pm0.008$	       \\
$l_3$               & $0.56$, $0.57$, $0.58$ ($\pm0.02$)    & $0.30$, $0.34$, $0.36$ ($\pm0.01$)    & $0.366 \pm0.003$        & $0.374 \pm0.004$        & $0.345 \pm0.003$        & $0.361 \pm0.003$	       \\
    \hline
    \end{tabular}
    \label{table_wd}
\end{table*}

Based on the RV results in Tables \ref{tab:rv_js_fit} and \ref{tab:rv_fit} in Section~\ref{section:spectrometry}, the mass ratio ($q$) was fixed at 0.473. Consequently, the input range of the surface potentials  ($\Omega_1$, $\Omega_2$) was set to 3.5 -- 5.0.  Taking into account the RV  and $O-C$ analyses, the input range for eccentricity ($e$) was set to 0.09 -- 0.15.  The orbital cycle number of NO Pup A corresponding to the start  and end times of the TESS data was calculated using the linear ephemeris given in Table \ref{tab:o-c_fit}, and the argument of periastron ($\omega$) corresponding to these cycle numbers were calculated from Equation \ref{eq:omega}. Accordingly, the input range for $\omega$ was entered as $100^\circ$ to $140^\circ$. 

A range from 0.3 to 0.8 was set for the fractional luminosity of the primary component ($L_1$).  {The input range for the third contribution of light to the total light of the system ($l_3$) was set to 0.30 to 0.50, based on the CROWDSAP parameter discussed above with the recognition that NO Pup is a multiple star.}

A quadratic limb-darkening law was assumed; the coefficients were taken from \citet{Claret_2017} according to the effective temperatures and the filter used. The bolometric gravity darkening exponents and albedoes for both components were taken as 1.0, assuming that the components have radiative atmospheres, in accordance with the regular procedure of the WD program.

The values of the adopted WD + MC model for all the sectors' data are listed in Table~\ref{table_wd}. A comparison of the sectors' LCs with the WD + MC fits is provided in Figure~\ref{fig_lc_wd}. 


\subsection{WD+MC fits to BVR Photometry}
\label{sec:wd_bvr_fits}

We also used the {\sc wd+mc} program for simultaneous fittings of our ground-based BVR data. The input ranges of the adjustable parameters ($e$, $\omega$, $i$, $T_2$, $\Omega_1$, $\Omega_2$, $L_1$, $L_2$ and $L_3$) were entered into the program as with the TESS data (Section \ref{sec:wd_fits}). For the periastron longitude parameter, $\omega$, 80 to 110-degree limits were entered, according to the observation epochs, calculated from Equation \ref{eq:omega}. 

The orbital eccentricity, $e$, was found to drop below the reasonable limit of 0.10 when allowed to be set by the optimiser program. We have associated this with the effects of data irregularities, especially around the secondary minimum. Subsequently $e$ was fixed at $e =  0.13$. 

The adopted parameters of the WD+MC model for the the BVR LCs are presented in Table~\ref{table_wd}. {However, when these results were compared 
with the TESS results given in 
Table!\ref{table_wd}, it transpired that allowing freely  adjustable third light contributions ($l_3$) could change their nominal values by a factor of up to 2.} In an alternative approach to the BVR fittings, the geometric parameters ($e$, $i$, $\Omega_1$ and $\Omega_2$) were fixed at those found in the TESS LC analyses and just $T_2$, $\omega$, phase shift, $L_1$, $L_2$ and $l_3$ were left free.

We have thus referred to the normal approach as producing Model I, and the alternative, with the geometric elements  taken as constant, as Model II. The results of Model I and  II fittings are given together in Table~\ref{table_wd}. The third light contribution ($l_3$) in Model II  is consistent with the $l_3$ values obtained in other solutions. A comparison of Model II with the WD+MC and BVR LCs is shown in Figure \ref{fig_bvr_wd}. 

{We assumed that the third light came from the B component of NO Pup and calculated the relative light contributions of each component to the total light of the system (as $l_1+l_2+l_3$) in each band at 0.25 phase from the Model II results in Table \ref{table_wd}. These $l_1$, $l_2$  and $l_3$ values, and their corresponding magnitudes and colours, are given in Table \ref{tab:WD_colors}.  In this calculation, the magnitude $V = 6.086$ and colour indices $B-V = -0.019$ and $V-R = -0.002$ given in Section \ref{sec:wf_bvr_fits} for NO Pup were used. The dereddened colour indices were obtained from the colour excess $E(B-V) = 0.020$ derived from the SED analysis in Section \ref{sec:absolute_parameters}. Spectral types correspond to the $(B-V)_0$ colour index according to the calibration in \citet{Budding_2022} and in \citet{2018MNRAS.479.5491E}.}

{When comparing the WinFitter results (Table \ref{tab:mark_c_separate}) with the WD+MC results (Table \ref{tab:WD_colors}) for the magnitudes and colours of the components, there are small discrepancies of a few percent within the error limits in the magnitudes and colours of the Aa and Ab components, while the slightly larger discrepancy in those of the B component, and thus its spectral type shifts slightly to the early A spectral type in the WD+MC estimation. However, since component B is probably a binary star (see Section \ref{sec:intro}), it does not seem possible to make a definitive estimate for this component.}

\begin{table}[!bt]
    \caption{{Magnitudes and colours of NO Pup Aab and B from WD+MC re\-s\-ult\-s. Errors are on the order of 0.02 mag. } }
    \centering
    \begin{tabular}{l|c|c|c}
    \hline
Parameter   & Aa & Ab   & B  \\
\hline
$l_{1,2,3}$ (B,V,R)   & 0.65, 0.61, 0.57 & 0.05, 0.06, 0.07 & 0.30, 0.34, 0.36 \\
$B$ & 6.536  & 9.39 & 7.36  \\
$V$ & 6.630  & 9.16 & 7.27  \\
$R$ & 6.692  & 8.98 & 7.21  \\
$B-V$ & --0.094  & 0.23 & 0.09  \\
$V-R$ & --0.062  & 0.18 & 0.06  \\
$(B-V)_0$ & --0.114  & 0.21 & 0.07  \\
Spectral type   & B8V   & A7V   & A3V   \\
\hline
    \end{tabular}
\label{tab:WD_colors}
\end{table}


\subsection{Times of Minima}
\label{subsec:ToMS}

We followed the methodology of \citet{Zasche_2009} to analyse times of minima (ToM)s, extending the analysis of \citet{Wolf_2008}, who analysed data up to HJD 2452284.261 covering 25 minima. 
Our analysis is based on 53 ToMs extending from HJD 2441351.7094 to 2460013.952957 with the inclusion of estimates from TESS data (20 ToMs), our BVR photometry (2 ToMs), and 6 ToMs from \citet{Kreiner_2004}. 

To find the parameters related to the apsidal motion using the ToMs of NO Pup A, the methods given by \citet{Gimenez&Garcia_Pelayo_1983} and \citet{Gimenez&Bastero_1995} were used. Accordingly, each observed ToM of the eclipsing binary star with apsidal motion can be represented by the equation: 

\begin{equation}
    T = T_0 + EP_s + \Delta \tau,
    \label{eq:tom}
\end{equation}
where $P_s$ is the sidereal period and is related to the anomalistic period $P_a$ and apsidal motion rate $\dot{\omega}$  by:

\begin{equation}
P_s = P_a \left(1 - \frac{\dot{\omega}}{360}\right),
\label{eq:PsPa}
\end{equation}
{where $P_s$ and $P_a$ is in days, $\dot{\omega}$ is in degrees per cycle, and 360 is the number of degrees in one cycle.}
$\Delta \tau$ in Equation \ref{eq:tom} is the term which shows that the primary and secondary ToMs shift periodically in antiphase about the linear ephemeris and as a function of the eccentricity $e$, argument of periastron $\omega$ and orbital inclination $i$ of the relative orbit of the binary star (see \citet{Gimenez&Bastero_1995}, their Eqs  (15-21)). 

We performed our O-C analysis 
under the assumption of apsidal motion, taking the initial values for $T_0$ and $P_s$ from \citet{Kreiner_2004} and the input values for other apsidal motion parameters from \citet{Wolf_2008}.  Our results from the best-fitting model are given in Table~\ref{tab:o-c_fit}, while Figure~\ref{fig:oc_fits} plots the model fit against the ToMs, and also the rescaled residuals.  Our findings are in good agreement with those of \citet{Wolf_2008}, with the apsidal period $U$ also in accord with that of \citet{Gimenez_1986}.
{However, although there are more data in our O-C analysis, the error estimates in the parameters of \citet{Wolf_2008} are generally smaller. A possible reason for this, apart from the different accuracies of the data samples used, is that different methods have been followed in the O-C analysis and therefore in the error estimation. We utilised a numerical method based on the incomplete gamma function for error estimation \citep[see][and his references]{2008arXiv0801.4258Z}, while \citet{Wolf_2008} use the least squares 
method to derive formal uncertainty values.
}

\begin{figure}[t]
\centering
	\includegraphics[scale=0.48]{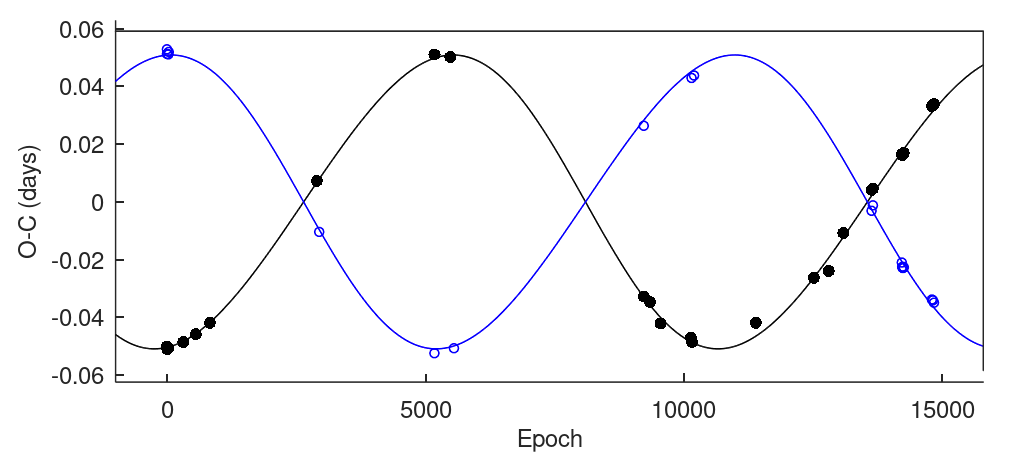} \\
    \includegraphics[scale=0.48]{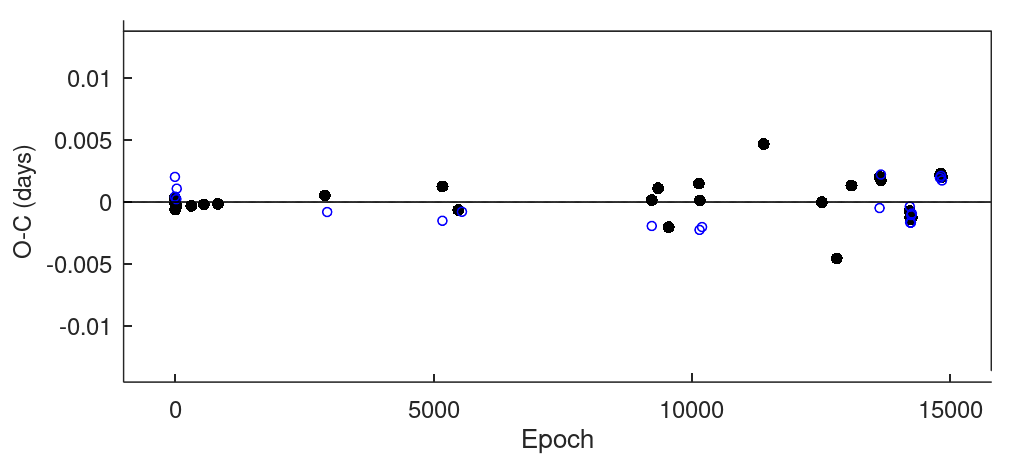}
\caption{In upper panel we plot the optimal model (shown as a black curve) for the primary minima (black filled circles), along with the optimal model fit to the secondary ToMs (blue curve and unfilled circles). Units are days. Lower panel shows the residuals from the optimal models.} 
\label{fig:oc_fits}
\end{figure}

\begin{table}[tb]
    \caption{Best fit estimates for the apsidal motion elements of NO Pup. See Figure~\ref{fig:oc_fits} for plots of the model fits to the times of minima. The parameter estimates from \citet{Wolf_2008} are given for easy reference.}
    \centering
    \resizebox{0.90\columnwidth}{!}{
    \begin{tabular}{l|c|c}
    \hline
Element                             & This study                            & \citet{Wolf_2008}
\\
\hline
$J\!D_0$ (HJD)                      & $ 2441361.8145 \pm 0.0094$            & $ 2441752.0741 \pm 0.0006$     \\
$P_o$ (days)                        & $ 1.2568804 \pm 0.0000009$            & $ 1.2568803 \pm 0.0000003$    \\
$e$                                 & $ 0.127 \pm 0.027 $                   & $ 0.1257 \pm 0.0012 $         \\
$\omega_0$ (deg)    & $ 2.9 \pm 2.6 $                       & $ 15.42 \pm 0.30 $             \\ 
$\dot{\omega}$ (deg/cycle)          & $ 0.033 \pm 0.001 $                   & {$ 0.03235 \pm 0.00025 $}        \\
$U$ (yrs)                           & $ 37.5 \pm 1.2 $                      & $ 38.29 \pm 0.04 $            \\
$P_a$ (days)                        & $ 1.2569956 \pm 0.0000009$           & $1.2569933 \pm 0.0000003$     \\
    \hline
    \end{tabular}}
    \label{tab:o-c_fit}
\end{table}

\begin{figure}[!t]
	\includegraphics[scale=0.43]{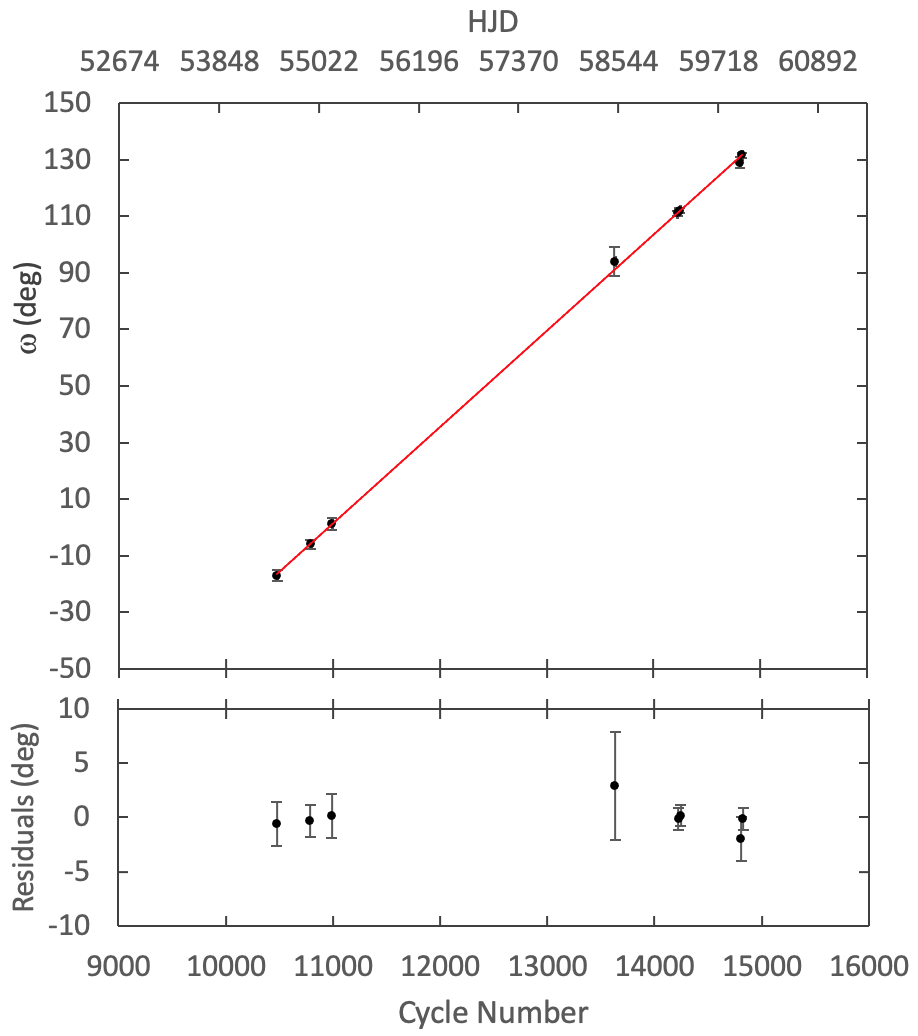}
\caption{Variation of the argument of periastron $\omega$ of the eccentric binary NO Pup A.} 
\label{fig_omega}
\end{figure}

Keeping in mind the variation of the argument of periastron ($\omega$) determined from the O-C analysis, as an alternative approach, we plotted the $\omega$ values found from the solutions of the RV and LCs of the system according to the cycle number in Figure \ref{fig_omega}.  The $\omega$ values in the lower left of the figure are taken from the analysis of the HARPS RV curves  (Table \ref{tab:rv_js_fit}), the RV1 solution from the UCMJO spectra and that from the REOSC spectra (Table~\ref{tab:rv_fit}), and the $\omega$ values in the upper right of the figure are from the BVR and TESS LC fittings (Table~\ref{table_wd}). In this way, we obtained the linear time dependence of $\omega$  shown in Figure~\ref{fig_omega}:
\begin{equation}
\label{eq:omega}
    \omega = \omega_0 + \dot{\omega} E,
\end{equation}
where $E$ is the cycle number of the mid-times of the related observations. These were calculated from the linear ephemeris given in Table \ref{tab:o-c_fit}. The plot of the $\omega$ values against cycle number in Figure \ref{fig_omega} reveals that $\dot{\omega}$ = $0.0340 \pm0.0003$ deg/cycle and $\omega_0$ = $-373 \pm4$ deg as the linear best fit parameters. This value of $\dot{\omega}$, which was confirmed by the O-C analysis (see Table \ref{tab:o-c_fit}), results in a value of $U = 36.4 \pm 0.3 $ yrs.  



\section{Spectrometry}
\label{section:spectrometry}

\subsection{HARPS spectra}
\label{subsection:harps_spectra}

{The European Southern Observatory (ESO) Science Archive Facility (SAF)\footnote{http://archive.eso.org/cms.html} contains 43 HARPS (High-Accuracy Radial-velocity Planet Searcher) spectra of NO Pup, taken between December 09, 1996 and June 21, 2015.\footnote{ESO proposal 083.D-0040(A), PI J.\ Southworth.} }  Of these spectra, we selected 34 from the nights of April 2 to 7, 2009. We used the HARPS cross-dispersed \'echelle spectrograph at the 3.6-m telescope at La Silla Observatory \citep{2003Msngr.114...20M}. We chose to operate HARPS in the EGGS mode, which has a larger fibre entrance on the sky than the standard HAM mode (1.4 versus 1.0 arcsec), resulting in a higher throughput (by a factor of approximately 1.75), a lower resolving power (80,000 as against 115,000) and a lower but still excellent RV precision (3\,m\,s$^{-1}$ rather than 1\,m\,s$^{-1}$).


Each spectrum consists of 72 orders incident on two CCDs that cover the range 3780--6900~\AA\ with a gap at 5304--5337~\AA\ between the CCDs. The data were reduced using the standard HARPS pipeline, with the pipeline products retrieved from the ESO SAF.


\subsection{FEROS spectra}
\label{subsection:feros_spectra}

Upon reviewing source material we found that the ESO SAF \footnote{http://archive.eso.org/cms.html} maintains several datasets from the Fiber-fed Extended Range Optical Spectrograph (FEROS) on the 2.2-m telescope at La Silla \citep{2012MNRAS.420.2727E}. This \'{e}chelle spectrograph has a resolving power of 48,000 and each exposure covers the full optical range (3600--9200\,\AA). The majority of these data were obtained under ESO Program ID 088.D-0080(B) (PI: He{\l}miniak). 
{However, since the HARPS spectra have a higher spectral resolution and signal-to-noise (S/N) ratio (on average 170), we preferred to use only the HARPS spectra in the present study. We discuss these data next. }


\subsection{{HARPS spectral analysis}}
\label{subsection:harps_spectra_analysis}

{Although all 43 HARPS spectra available at ESO SAF were taken into account for RV measurements and  atmosphere modelling (see Section \ref{sect:atm}), the consequences  of the rapid apsidal motion caused us to select only RVs from the 34 HARPS spectra during a 5-day  interval for the spectroscopic orbit model. }

%

{For the RV measurements}, we employed the cross correlation technique using the IRAF FXCOR task \citep{1986SPIE..627..733T}. As template material, we generated synthetic spectra based on the known spectral types of the binary components (Sections \ref{sec:intro} and \ref{sec:photometry}). {For each binary component, two different theoretical templates were generated based on their spectral types and the spectral type–effective temperature relationship \citep{2000asqu.book.....C}, using ATLAS9 model atmospheres \citep{1993KurCD..13.....K} and the {\sc synthe} code \citep{1981SAOSR.391.....K} as the two components have significantly different spectral types. These synthetic templates were then broadened according to the spectral type–projected rotational velocity relationship \citep{2000asqu.book.....C}.} RV variations were determined and are listed in Table \ref{tab:rv_harps}. {These HARPS RVs were phased according to the linear ephemeris given in Table \ref{tab:o-c_fit} }. Additional spectral analysis to estimate atmospheric parameters is provided in Section~\ref{sect:atm}.

{The selected 34 HARPS RV were fitted using the DC mode of the WD program. } The semi-major axis ($a$), eccentricity ($e$) and periastron longitude ($\omega$) of the relative orbital ellipse, the phase-shift ($\Delta \phi$), systemic velocity ($\gamma$), and mass ratio ($q$) of the binary system were taken as adjustable parameters.
{However, due to the orbital eccentricity tending to drop below the assigned limit of 0.10, presumably due to insufficient optimal parameter resolution, $e$ was fixed to the weighted average of the values in the previous RV and LC solutions, i.e.\ 0.127. } 
The optimal values are given in Table \ref{tab:rv_js_fit}, and a comparison of the model with the observations is shown in Figure \ref{fig_js_rvc}. The amplitudes ($K_1$ and $K_2$) were calculated accordingly.  Similar results were found using the {\sc Winfitter} program. 

\begin{figure}
\centering
    \includegraphics[scale=0.48]{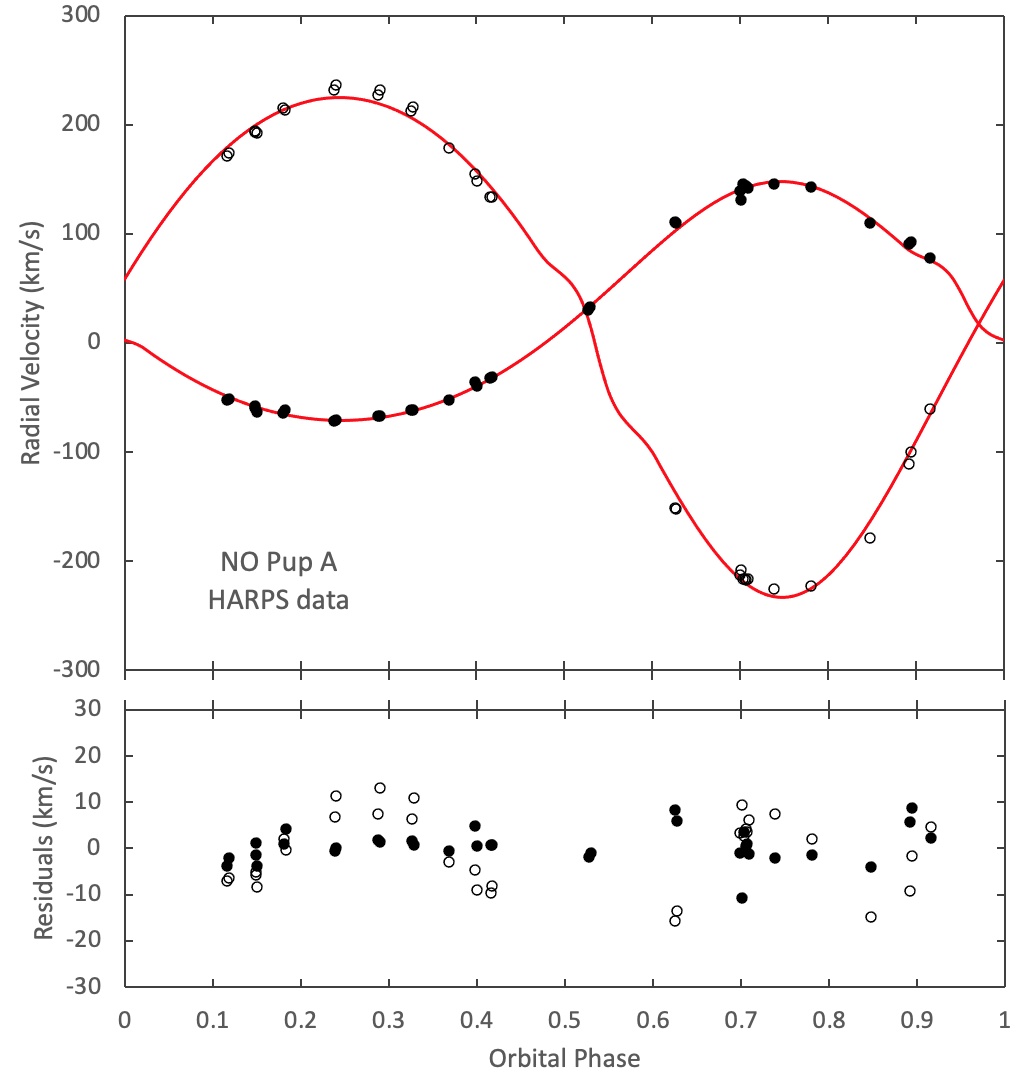}
    \caption{RV curves measured from selected HARPS spectra of NO Pup A with the WD model fitting. Residuals to the RV models are plotted in the bottom figure. RVs of the primary and the secondary components are marked as filled and hollow symbols, respectively.
    } 
    \label{fig_js_rvc}
\end{figure}

\begin{table}
    \caption{Best fit {\color{blue} WD} modelling for the RV curves measured from selected HARPS spectra of NO Pup A. }
    \centering
    \begin{tabular}{l|c}
    \hline
Parameter                       & Value   \\
\hline
Phase shift                     & $ -0.0042 \pm 0.0012$    \\
$a$~(R$_{\odot}$)               & $ 8.53 \pm 0.05$    \\
$e$                             & $ 0.127 $ (fixed)   \\
$\omega$ (deg)                  & $ 359 \pm 2 $   \\ 
$ \gamma $ (km s$^{-1}$)        & $ 24.7 \pm 0.8$     \\ 
$q=M_{2}/M_{1}$                 & $ 0.473 \pm 0.006 $   \\
\hline
$K_1$ (km s$^{-1}$ )            & $ 109.9 \pm 0.9$    \\
$K_2$ (km s$^{-1}$ )            & $ 232.1 \pm 1.5$  \\
\hline
    \end{tabular}
\label{tab:rv_js_fit}
\end{table}


\subsection{UCMJO spectra}
\label{subsection:ucmjo_spectra}

Relevant spectroscopic observations include those made with the HERCULES spectrograph  \citep{Hearnshaw_2003}, together with the 1m McLellan telescope at the University of Canterbury Mt John Observatory (UCMJO).

Observations were recorded with a 4k$\times$4k Spectral Instruments (SITe) camera \citep{Skuljan_2004}.
Wavelength and relative flux calibration was performed using the latest version of the software package {\sc hrsp} \citep{Skuljan_2004b, Skuljan_2021} that outputs measurable data in {\sc fits} \citep{Wells_1981} formatted files. Typical exposures lasted for  $\sim$700 seconds.  Further information on the spectroscopic arrangements at UCMJO were given by \citet{Budding_2024}.

24 exposures of NO Pup were made during the interval December 3 - 13, 2009, although,  unfortunately, a number of them were affected by technical difficulties. 
However, 10 spectra were selected from these spectral images for analysis. The individual lines detected in the orders in the observed spectra are listed in Table \ref{tab:spectral_features}. Since the light contribution of the secondary component is relatively low -- on the order of $\sim$10 percent (see Section \ref{sec:photometry}), mostly only the primary features could be distinguished.

Apart from  H$\alpha$ and H$\beta$, the best-defined line is probably the primary He I $\lambda$6678, where there is no significant contribution from the secondary since its effective temperature is  $\sim$7500 K (see Table~4). 
The He I $\lambda$6678 lines in these 10 spectra were studied, and the RVs of the primary determined, using the program {\sc Prof} \citep[latest version,][]{Erdem_2022}.

If the resolution is sufficiently high, {\sc Prof} models spectral line profiles with a parameter set that determines the RV of the centre of light, as well as the rotation rate of the source and the turbulence scale in the surrounding plasma. The application of {\sc Prof} to the He I $\lambda$6678 features resulted in the values of RV, rotation parameter ($r$) and equivalent width (EW) listed in Table \ref{tab:rv_ucmjo}, and an example of such a profile fitting is displayed in Figure \ref{fig:prof}.

The mean value of the rotation parameter ($r$) for the data sets in Table \ref{tab:rv_ucmjo} yields a projected mean equatorial rotation speed of 
{$82 \pm2$ km s$^{-1}$.} 
If we assume that the inclination of the primary star's rotation axis is equal to that of the system's orbit ($i_{rot}=i_{orb}$) and 
{we neglect effects related to the low orbital eccentricity}, 
the primary exhibits a synchronised rotation with the mean orbital revolution ($P_{rot}=P_{orb}$), we find the synchronous projected rotation speed of the primary as { $86 \pm5$ km s$^{-1}$ } 
using the formula $v_{rot}\sin i_{rot}$ = $(2\pi R_{1}\sin i_{rot})/P_{rot}$ and the values of $i$ and $R_1$ in Table \ref{table:abs_par}. In this way, the projected rotational velocity of the primary component, derived from the He I $\lambda$6678 line profile fitting, and the projected synchronous rotational velocity computed from Table~\ref{table:abs_par} agree within the uncertainty limits, supporting that NO Pup Aa rotates synchronously. {The possibility of pseudo-synchronous rotation in the NO Pup A system is considered in Section~\ref{sec8}. }




The fitting function in {\sc Prof} has two main components: uniform rotation and a Gaussian turbulence broadening. Here {\sc Prof} applied to the He I line in the UCMJO spectra of NO Pup A gives the turbulence parameter of the order of a few km s$^{-1}$ for the surface of the primary star.

{\sc Prof} also estimates the equivalent width of a line by numerical integration. The results are given in Table \ref{tab:rv_ucmjo}. The mean value of the measured equivalent widths (EWs) for the data sets in Table \ref{tab:rv_ucmjo} is { $0.07 \pm0.01$ \VB{\text{\AA}}. } 
The relative noise in the He I lines in the observed spectra makes for a fairly uncertain mean value of the EWs.  However, in comparison with the calibration data of \citet{Leone_andLanzafame_1998}, the effective temperature of the primary was estimated to be $12000 \pm1000$ K, which corresponds to a spectral type B8/9.

The UCMJO observations, adopted as of suitable quality,  provide coverage for the first half of the full radial velocity (RV) cycle, consistent with the spectroscopic data of \citet{Veramendi_2014}. Their observations were made with the 2.15 m telescope and the REOSC \'{e}chelle spectrograph at the Complejo Astronómico El Leoncito (CASLEO) during 10 allocations between 2008 and 2013.  RVs were determined  by the cross-correlation technique, with spectral disentangling of double-lined systems.

The spectroscopic results, as a whole, have confirmed the type classifications for NO Pup and obtained good orbital coverage, with twenty individual points having signal/noise ratios of greater than 100. \citet{Veramendi_2014}  went on to provide absolute parameters of the system, making use of \citet{Gronbech_1976}'s {\it uvby}  photometry and the well-known Russell paradigm.

\begin{table}[bt]
    \caption{Values of RV, rotation parameter ($r$) and equivalent width (EW) of the primary component of NO Pup A derived from the He I lines in the UCMJO spectra.} 
    \centering
    \begin{tabular}{lccccc}
    \hline
BJD             & Orbital   & RV1           & Err.          & $r$       & EW   \\         
2400000+        & phase     & km s$^{-1}$   & km s$^{-1}$   & km s$^{-1}$   & \VB{\text{\AA}}    \\
\hline
55170.0967	&	0.1554	&	-60.1	&	3.8	&	88	&	0.077	\\
55170.1414	&	0.1910	&	-62.1	&	4.5	&	84	&	0.080	\\
55172.8736	&	0.3647	&	-48.6	&	4.8	&	91	&	0.069	\\
55172.9421	&	0.4192	&	-36.4	&	5.4	&	76	&	0.058	\\
55177.9041	&	0.3671	&	-57.4	&	4.3	&	90	&	0.093	\\
55178.9096	&	0.1672	&	-67.3	&	4.1	&	91	&	0.100	\\
55178.9452	&	0.1954	&	-69.3	&	6.2	&	77	&	0.050	\\
55178.9834	&	0.2259	&	-69.9	&	5.0	&	80	&	0.062	\\
55179.0374	&	0.2688	&	-73.4	&	4.5	&	71	&	0.053	\\
55179.0939	&	0.3138	&	-67.0	&	3.9	&	76	&	0.084	\\
\hline
    \end{tabular}
    \label{tab:rv_ucmjo}
\end{table}

\begin{figure}[!bt]
\centering
	\includegraphics[scale=0.48]{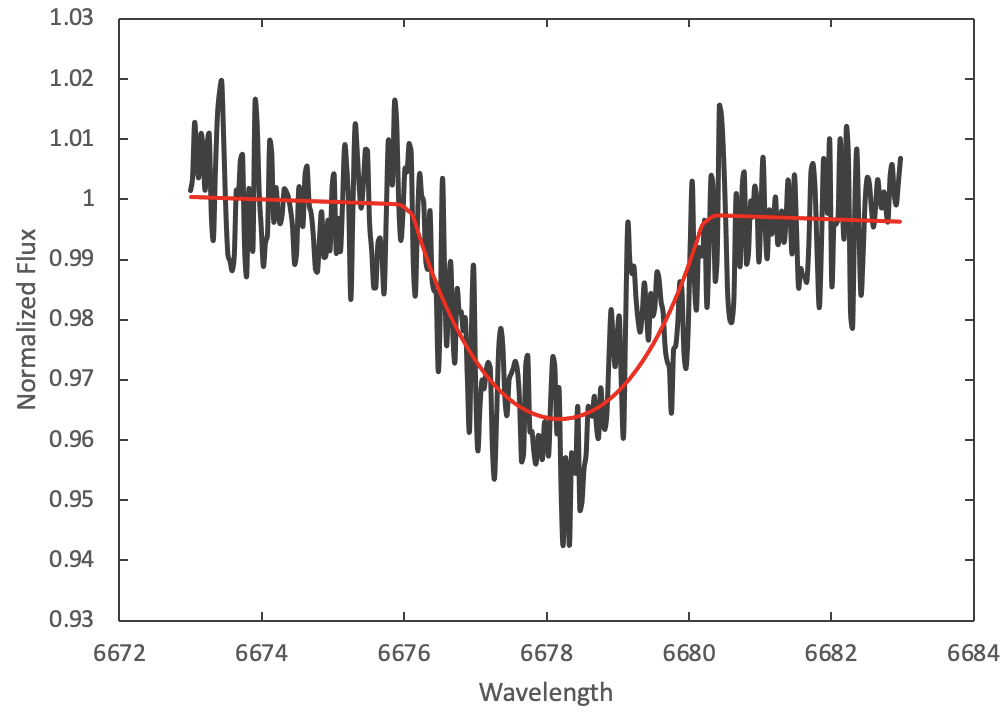}
\caption{Convolved rotation Gaussian fitting to the He I $\lambda$6678 line profile in UCMJO spectrum of NO Pup.} 
\label{fig:prof}
\end{figure}

\begin{table*}[bt]
    \caption{Best fit modelling  for the RV curves of \citet{Veramendi_2014} of NO Pup. 
    }
    \centering
    \begin{tabular}{l|c|c|c}
    \hline
Parameter                       & {\sc Winfitter} Model  & WD Model        & V \&  G (2014)                 \\
\hline
$P$                             & 1.256879          & 1.2568804            & 1.25700056                     \\
$T_0$                           & 2441361.815      & 2441361.8145         & 2441351.7568                   \\
$T_{\rm ref}$                   &  2454538.119      & 2454544.272                  &  ---                           \\
$K_1$ (km s$^{-1}$ )            & $ 107.7 \pm 2.0$  & $ 106.9 \pm 1.4$     & $ 107.5 \pm 1.8 $              \\
$K_2$ (km s$^{-1}$ )            & $ 222.0 \pm 2.0$  & $ 224.1 \pm 2.0$     & $ 226.2 \pm 3.0 $              \\
${\Delta} {\theta_0}_0$ (deg)   & $ 14.4 \pm 0.4 $  & ---                  &  ---                           \\
Phase shift                     & ---               & 0.0024               & ---                            \\
$ \gamma $ (km s$^{-1}$)        & $ 25.2 \pm 1.3$   & $ 24.8 \pm 0.7$      & $25.4\pm 0.6$                  \\ 
$e$                             & $0.127 \pm 0.027$ & $ 0.123 \pm 0.006 $  & $ 0.1249 \pm 0.0004 $          \\
$M_0$ (deg)                     & $ 98.2 \pm 16 $   & ---                  &   ---                          \\ 
$\omega$ (deg)                  & $ 338 \pm 16$     & $ 343.6\pm 2.5 $       & $ 6.1\pm 1.2 $                 \\ 
    \hline
    \end{tabular}
    \label{tab:rv_fit}
\end{table*}

\begin{figure}[!t]
\centering
	\includegraphics[scale=0.48]{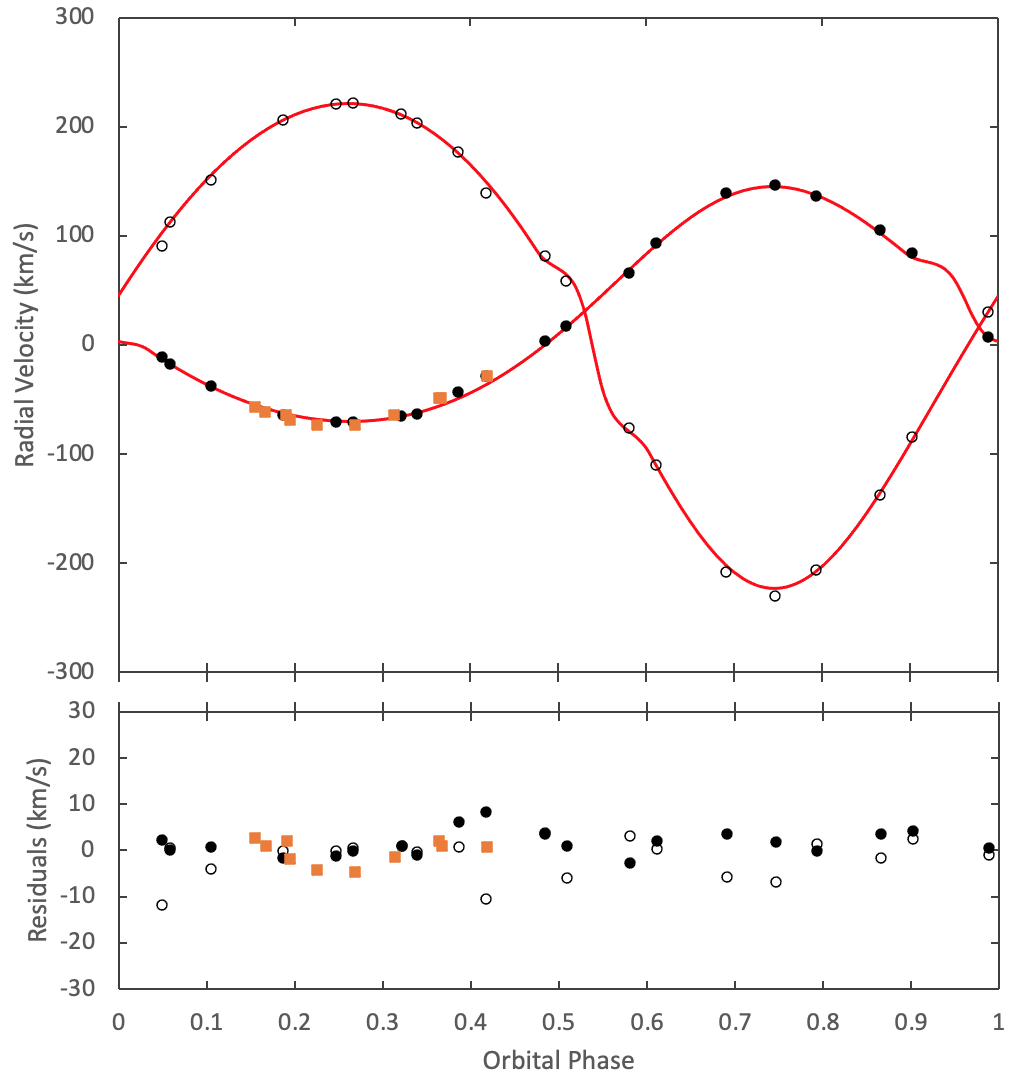}
\caption{
RV curves of NO Pup A with the WD model fitting. Black circles denote REOSC RV data of \citet{Veramendi_2014}, whereas orange circles denote the RVs of the primary component derived from He I lines in the MJUCO spectra. 
Residuals to the RV models are plotted in the bottom figure. RVs of the primary and the secondary components are marked as filled and hollow symbols, respectively.
} 
\label{fig_rvc}
\end{figure}

We remodelled the RV data set of \citet{Veramendi_2014} using the program suites {\sc Winfitter} \citep{Rhodes_2023} and {\sc WD} and presented our results in Table~\ref{tab:rv_fit}. 
{Although the main parameters of the optimal RV curve fittings are essentially similar to those of \citet{Veramendi_2014}, the values of the longitude of the periastron ($\omega$) appear  different. Noting the decline in precision of the determination of $\omega$ at low $e$, the estimates of {\sc Winfitter} and  {\sc WD}  are tolerably
in agreement; however, the $\omega$ value of \citet{Veramendi_2014} is quite closer to zero. This may reflect a different reference epoch for the  data of \citet{Veramendi_2014}, given the fast rate of apsidal motion ($\sim$10 deg y$^{ -1}$).}


\section{Determination of the atmospheric parameters} \label{sect:atm}

The atmospheric parameters of a star, in particular the effective temperature (\teff) and surface gravity (\logg), are crucial for understanding the nature of the source. This is especially true for binary stars, where accurately estimating the \teff\, values of both components is essential for performing reliable binary modelling analysis. 

There are two methods that can be applied in advance to determine the atmospheric parameters of binary systems. The first involves creating composite synthetic spectra and comparing them with the observed binary spectra. The second method is spectral disentangling, which allows us to separate the spectra of both binary components. For both methods, the flux ratios of the binary components must be known. However, for spectral disentangling, reliable results require spectra spread over the orbital phase. Since our data have such a distribution, we preferred applying the spectral disentangling.  This is explained more fully in the next section.



\subsection{Disentangling  Components' Spectra and Atmosphere Modelling}
\label{subsec:atmospheres}

\begin{figure*}[!t]
\centering
\includegraphics[width=0.35\columnwidth]{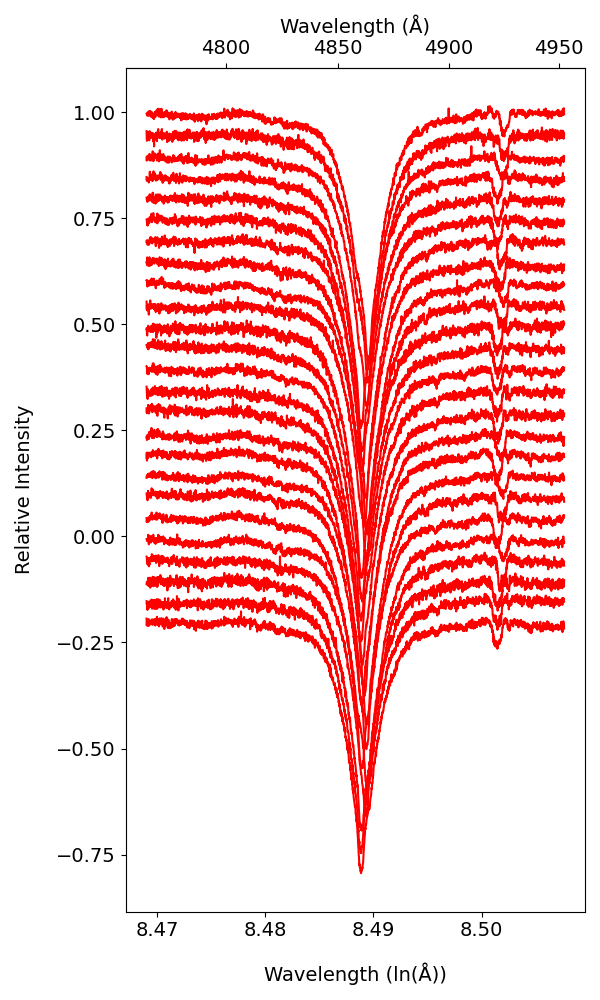} 
\includegraphics[width=0.35\columnwidth]{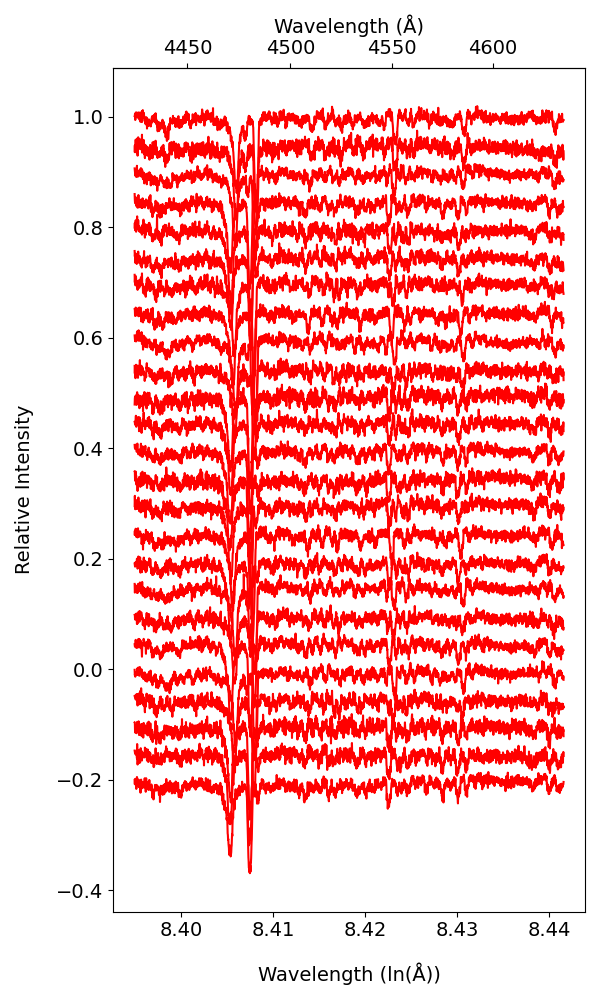} \\
\caption{Spectral regions around H$\beta$ in the HARPS data used for disentangling. \label{fig:Hbeta_reg}}
\end{figure*}

We have selected two spectral regions in the {\sc HARPS} data to disentangle the component spectra. Both regions are centred around the H$\beta$ line, as shown in Figure~\ref{fig:Hbeta_reg}. This region was chosen due to the presence of a relatively high number of metal lines for both components.  This enables the extraction of their individual spectra and facilitates the modelling of the spectroscopic orbit. The codes {\sc korel} \citep{2004PAICz..92...15H} and {\sc fdbinary} \citep{2004ASPC..318..111I} were used to cross-check the results.

Spectra obtained during eclipses, except at mid-eclipse, were excluded from the analysis, because non-Keplerian effects, such as the Rossiter-McLaughlin effect, are not accounted for in the codes. In the {\sc korel} analysis, each spectral region was re-binned to 2048 bins, resulting in a resolution of approximately 7~km~s$^{-1}$ per pixel. This re-binning smoothed the data without significantly compromising resolution. For the {\sc fdbinary} analysis, the spectra were used directly, with no additional processing applied before the disentangling. During the analysis, some orbital parameters, such as $P$ and $K_{1,2}$ were held constant. Additionally, the flux ratio of the binary components was examined.

The disentangled spectra determined from both {\sc korel} and {\sc fdbinary} programs were modelled using synthetic spectra computed from Kurucz' {\sc atlas9} model atmosphere grids \citep{1993KurCD..13.....K}. In the atmospheric determination with the {\sc korel} disentangled spectra, the grid parameters and intervals for the primary and secondary components were chosen as given below. During this analysis, the microturbulence parameter, 
{$\xi$}, 
was set at 2 km/s for both components.

\begin{itemize}
    \item \textbf{Primary:}
    \begin{itemize}
        \item Effective temperature ($T_\mathrm{eff}$): 11,000--14,000~K, in steps of 100~K
        \item Surface gravity (log $g$): 4.3--4.4~cgs, in steps of 0.05~cgs
        \item Metallicity ([M/H]): $-0.5$ to $+0.5$, in steps of 0.5
        \item Projected rotational velocity ($v \sin i$): 60--100~km~s$^{-1}$, in steps of 10~km~s$^{-1}$
    \end{itemize}
    \item \textbf{Secondary:}
    \begin{itemize}
        \item $T_\mathrm{eff}$: 7000--9000~K, in steps of 100~K
        \item Surface gravity (log $g$): 4.3--4.4~cgs, in steps of 0.05~cgs
        \item Metallicity ([M/H]): $-0.5$ to $+0.5$, in steps of 0.5
        \item $v \sin i$: 60--100~km~s$^{-1}$, in steps of 10~km~s$^{-1}$
    \end{itemize}
\end{itemize}
\noindent
with  {\sc fdbinary}. 
the input parameters used in the spectral analysis of the separated spectra were selected a follows:
\begin{itemize}
    \item \textbf{Primary:}
    \begin{itemize}
        \item Effective temperature ($T_\mathrm{eff}$): 10000--14000~K, in steps of 100~K
        \item Surface gravity (log $g$): 3.8--4.4~cgs, in steps of 0.1~cgs
        \item Metallicity ([Fe/H]) : $-0.5$ to $+0.5$, in steps of 0.1
        \item $v \sin i$: 30--150~km~s$^{-1}$, in steps of 1~km~s$^{-1}$
    \end{itemize}
    \item \textbf{Secondary:}
    \begin{itemize}
        \item $T_\mathrm{eff}$: 6800--8500~K, in steps of 100~K
        \item Surface gravity (log $g$): 3.8--4.4~cgs, in steps of 0.1~cgs
        \item Metallicity ([Fe/H]): $-0.5$ to $+0.5$, in steps of 0.1
        \item $v \sin i$: 30--150~km~s$^{-1}$, in steps of 1~km~s$^{-1}$.
    \end{itemize}
\end{itemize}

In this analysis, the spectrum synthesis method \citep{2006ESASP.624E.120N} was used to estimate the atmospheric parameters ($T_\mathrm{eff}$, log $g$, $\xi$) and iron (Fe) abundances of both close binary components by taking into account the Kurucz line list\footnote{kurucz.harvard.edu/linelists.html}. Based on the Saha-Boltzmann equation, the atmospheric parameters were estimated during the analysis in the way presented by \citet{2016MNRAS.458.2307K}.

The resulting atmospheric parameters, after examination of the \texttt{KOREL} and \texttt{FDBINARY} disentangled spectra, are given in Table\,\ref{tab:atmospar}. As can be seen in the table, the results of both analyses agree within the adopted error limits. 

{Our results are consistent with those of \citet{Veramendi_2014} within the error bars; however, the \vsini\, of the secondary component shows a slight discrepancy (58.7\,$\pm$ 0.6\,km~s$^{-1}$ from \citeauthor{Veramendi_2014}, \citeyear{Veramendi_2014}). Given that the secondary component is fainter and the spectral data used by \citet{Veramendi_2014} have lower resolving power and signal-to-noise ratio, we consider this difference acceptable.} 
The disentangled spectra of both components, along with the best-fitting synthetic spectra, are displayed in Figure\,\ref{fig:metal_dis} and Figure\,\ref{fig:Hbeta_dis}. 
{The derived  effective temperatures in Table~\ref{tab:atmospar} appear slightly greater than those  given in Table~\ref{table_wd}, but the differences are within reasonable uncertainty estimates of each other.}

\begin{table}[!t]
\centering
  \caption{The results of atmospheric parameter analyses for the primary and secondary components of No\,PUP using the \texttt{KOREL} and \texttt{FDBINARY} disentangled spectra. The metallicity values given in the table are the [M/H] and [Fe/H] for the \texttt{KOREL} and \texttt{FDBINARY} analyses, respectively. * represents the fixed parameters.}
  \label{tab:atmospar}
\begin{tabular}{lcc}
\hline
Parameters                      & \texttt{KOREL}        &  \texttt{FDBINARY}    \\
\hline
                                  &\multicolumn{2}{c}{Primary}\\  \hhline{~--}
$T_\mathrm{eff}$ (K)            &13700\,$\pm$\,300      &13500\,$\pm$\,300     \\
log $g$ (cgs)                   &4.39\,$\pm$\,0.05      &4.3\,$\pm$\,0.1        \\
$\xi$ (km\,s$^{-1}$)            &2*                     &2.4\,$\pm$\,0.2        \\
$v \sin i$ (km\,s$^{-1}$)       &85\,$\pm$\,5           &86\,$\pm$\,2           \\
Metallicity                     &0.0\,$\pm$\,0.25       &$-0.02$\,$\pm$\,0.01     \\
                                &\multicolumn{2}{c}{Secondary}\\ \hhline{~--}
$T_\mathrm{eff}$ (K)            &8100\,$\pm$\,1000     &7500\,$\pm$\,200       \\
log $g$ (cgs)                   &4.35\,$\pm$\,0.05      &4.2\,$\pm$\,0.1        \\
$\xi$ (km\,s$^{-1}$)            &2*                     &2.1\,$\pm$\,0.2        \\
$v \sin i$ (km\,s$^{-1}$)       &65\,$\pm$\,5           &64\,$\pm$\,2           \\
Metallicity                     &0.0\,$\pm$\,0.25       & $-0.08$\,$\pm$\,0.03    \\
\hline
\end{tabular}
\end{table}




\begin{figure}
\centering
\includegraphics[width=1.0\columnwidth]{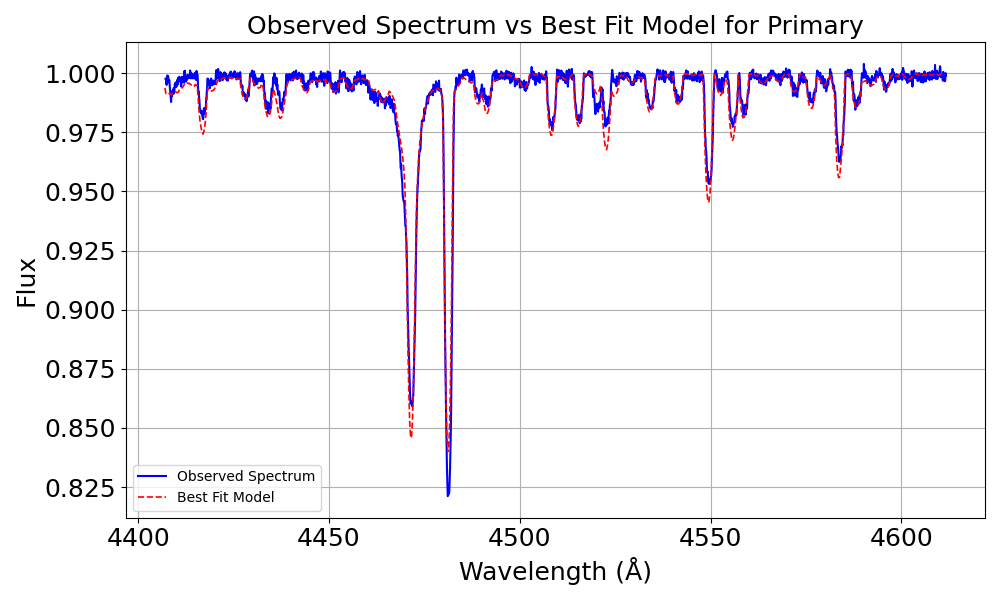} \\
\includegraphics[width=1.0\columnwidth]{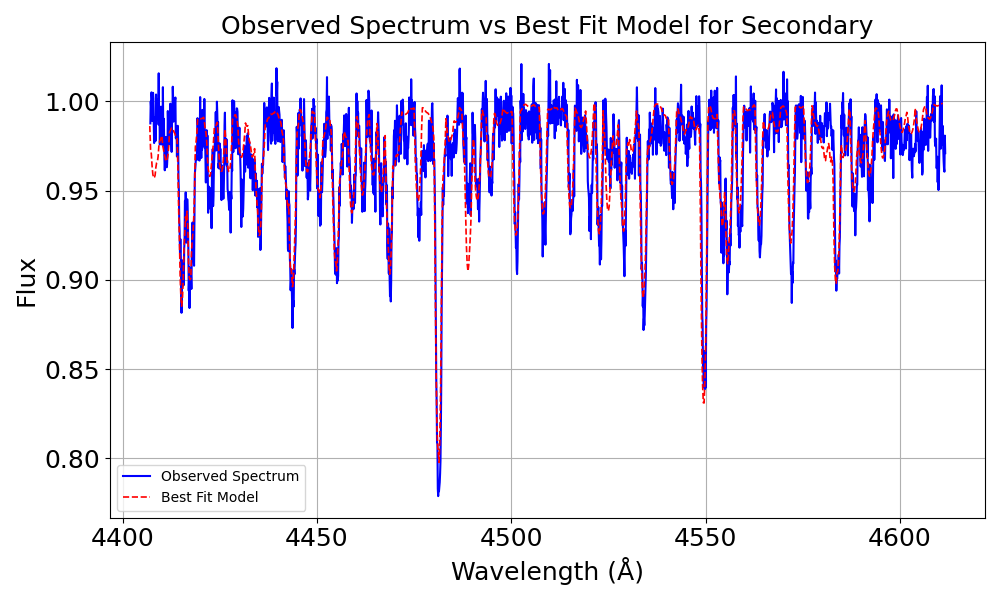}
\caption{Disentangled metal lines and best fitting synthetic spectrum for the primary (top) and secondary (bottom) components, respectively. \label{fig:metal_dis}}
\end{figure}


\begin{figure}
\centering
\includegraphics[width=0.8\columnwidth]{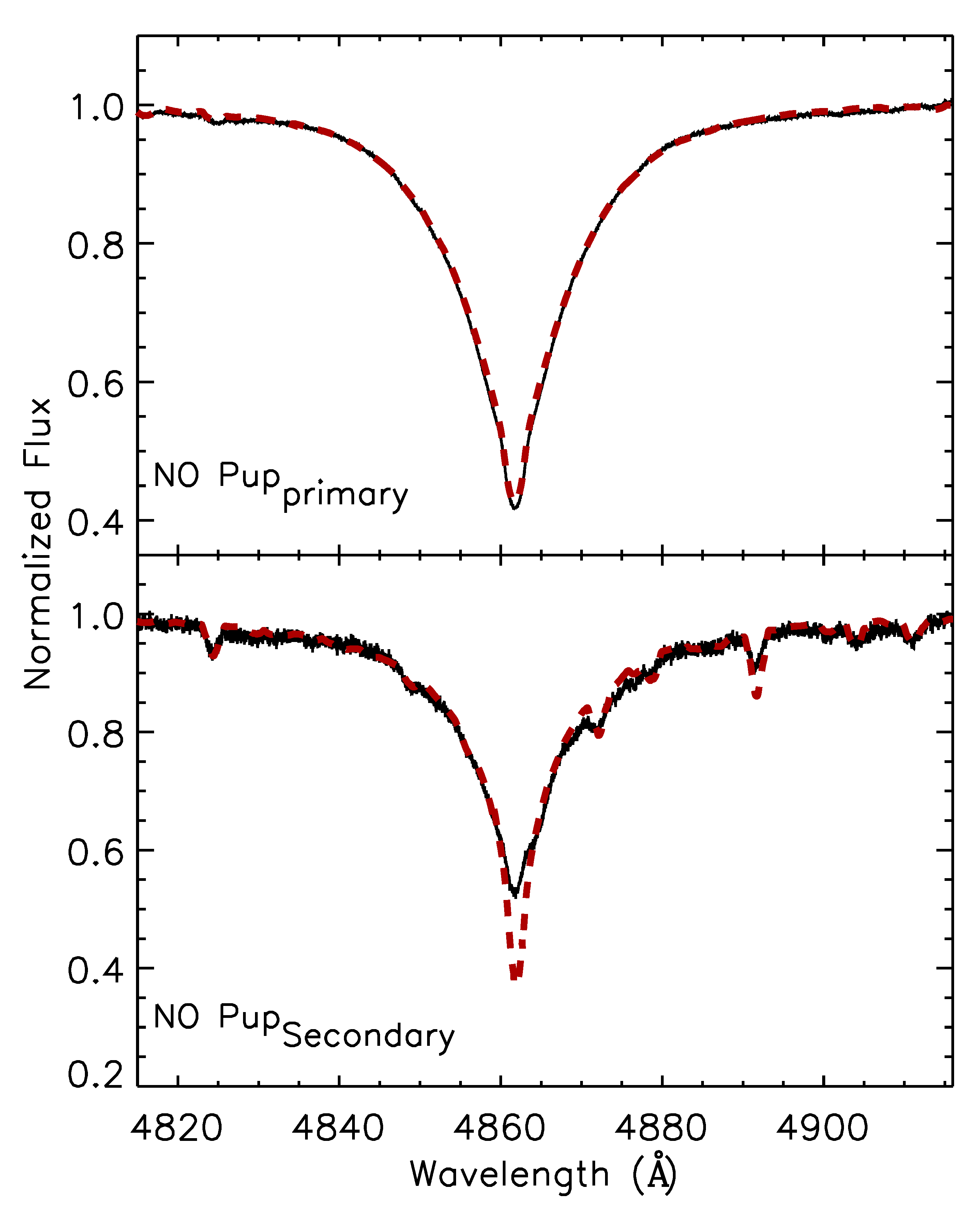}
\caption{Disentangled H$_\beta$ line and best fitting synthetic spectrum for the primary (top) and secondary (bottom) components, respectively. \label{fig:Hbeta_dis}}
\end{figure}


\section{Absolute parameters}
\label{sec:absolute_parameters}

Applying the well-known rearrangement of Kepler's third law as:  
\begin{equation}
\label{kepler}
    (M_1 + M_2) \sin^3 i = CP (1 - e^2)^{3/2} (K_1 + K_2) ^3 ,
\end{equation}
where the constant $C = 1.03615 \times 10^{-7} $, period $P$ is in days, and the RV amplitudes $K_1$ and $K_2$ are in km s$^{-1}$. 
{Adopting the weighted average values of eccentricity and orbital inclination from Section \ref{sec:photometry} as $e = 0.127$ and $i = 81.33^\circ$ and taking the RV amplitudes $K_1$ and $K_2$ from Table \ref{tab:rv_js_fit}, we find the component masses of NO Pup A to be $M_1 = 3.58 \pm 0.11$ and $M_2 = 1.68 \pm 0.09$ in solar units.}
These masses are slightly larger than those found using the RV data of \citet{Veramendi_2014} in Table \ref{tab:rv_fit}, 
perhaps due to the larger values of $K_1$ and $K_2 $ from the HARPS analysis. 

With the total mass of the close binary at {$5.26 \pm 0.20$ M$_{\odot}$} 
and the period 0.0034411 yr, Kepler's third law yields the semi-major axis of the orbit 
as {$0.03956 \pm 0.00023$ AU, or $8.51 \pm 0.05$ solar radii}. 
{From the results given in Section \ref{sec:photometry} we find $R_1 = 2.17 \pm 0.03$ and $R_2 = 1.51 \pm 0.06$.}

The high accuracy of this determination allows us to constrain the temperatures used in the photometric parallax (see Section \ref{sec:wd_fits} and the following paragraphs). 
{The  mean effective temperatures,
taking into account the results given in Sections \ref{sec:photometry}-\ref{sect:atm},
are, for the primary, $\sim 13300 \pm 500$; and 
for the secondary {$\sim 7400 \pm 500$ K}.
}


The surface gravitational accelerations ($g_1$, $g_2$) are directly related to solar values through $g/g_{\odot}= (M/M_{\odot}) / (R/R_{\odot})^2$.  The bolometric magnitudes ($M_{bol}$) and luminosities ($L$) of the component stars are calculated using Pogson's formula and the absolute radii and effective temperatures listed in Table~\ref{table:abs_par}.
We thus write $M_{bol}=M_{bol,\odot}+10\log T_{\odot} - 10\log T - 5\log (R/R_{\odot})$, and $L/L_{\odot}=10^{0.4(M_{bol,\odot}-M_{bol})}$.  The solar values, adopted by IAU 2015 Resolutions B2 and B3, were used in our calculations.  

{Bolometric corrections in the V band for the components ($BC_{1,2}$) were taken from \citet{Flower_1996}, while those in the TESS band were taken from \citet{2023MNRAS.523.2440E}, according to their effective temperatures used in the conversion from bolometric magnitudes to V and TESS-band absolute magnitudes ($M_{V,1,2}$ and $M_{TESS,1,2}$). 
The absolute magnitude of the eclipsing binary NO Pup A is also computed from following equation:}

%
\begin{equation}
    M_{band, \text{system}}=M_{band, 2} -2.5 \log \left(1+10^{-0.4\left(M_{band, 1}-M_{band, 2}\right)}\right).
\end{equation}
Finally, the distance to the system is calculated from the distance modulus ($d = 10^{m_{band} - M_{band} + 5 - A_{band}}$). Here $m_{band}$ is the apparent magnitude and $A_{band}$ is the interstellar extinction in the given band. 

The $A_{band}$ extinction is estimated from the SED analysis following the method described in \citet{Bakis_Eker_2022}, later refined by \citet{2023MNRAS.523.2440E} for the TESS pass-band. The best-fitting SED model determines the reddening as E(B--V)=0.02$\pm$0.01 mag, corresponding to $A_V$=0.062$\pm$0.031 mag and $A_{TESS}$=0.041$\pm$0.019 mag. Figure~\ref{fig:SED} presents the SED data alongside synthetic spectra computed using the system parameters, demonstrating a strong agreement between the model and observations. The absolute parameters thus obtained are given in Table \ref{table:abs_par} with their errors.

\begin{figure}[!tb]
\centering
\includegraphics[width=0.99\columnwidth]{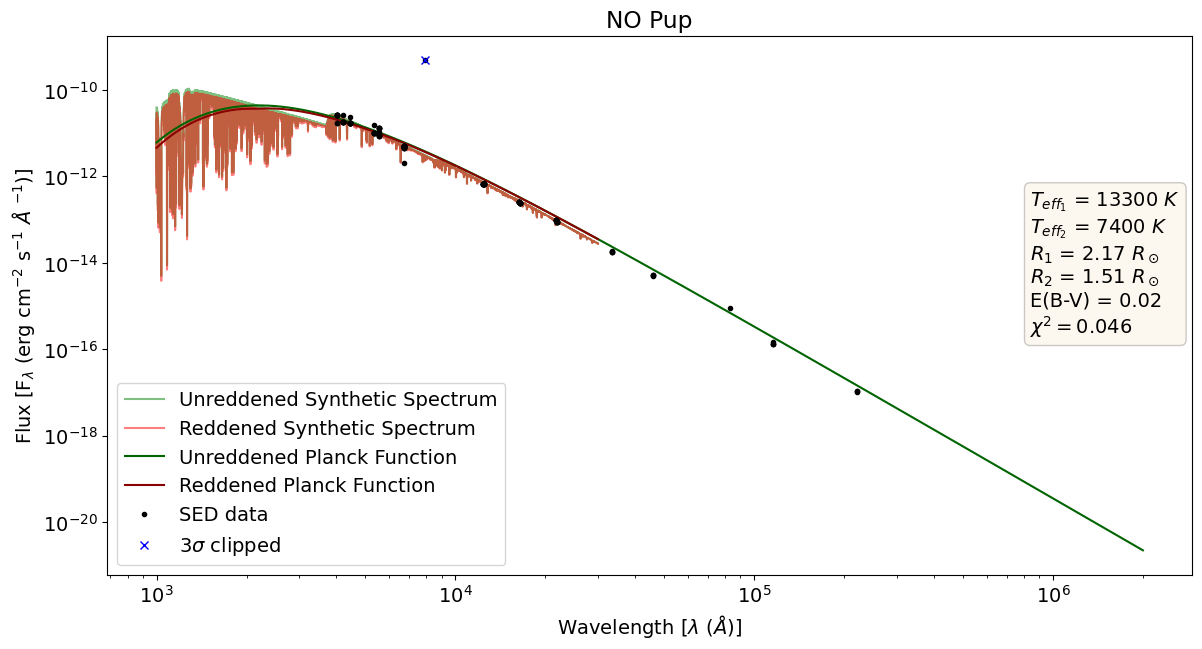}
\caption{SED data (black dots) and the combined synthetic spectra of the components, which are calculated using the absolute parameters of the components and the distance of the system given in Table~\ref{table:abs_par}. \label{fig:SED}}
\end{figure}

{On the other hand, the following equations given by \citet{Budding_Demircan_2007} can also be used for the photometric parallax:}
\begin{equation}
\log \Pi  = 7.454 - \log R - 0.2 \text{V} - 2 F'_\text{V} \\
\label{eq:photo_paralax}
\end{equation}
{where 
$F'_\text{V}$ (flux scale) is equal to 0.25 $\times$ the logarithm of the surface flux in the V band, and is specified by}
\begin{equation}
F'_\text{V} = \log T_{eff} + 0.1 BC
\label{eq:flux_scale}
\end{equation}
{Here, if we use the equations given above for the NO Pup Aa component, i.e. $V = 6.630$ from Table \ref{tab:WD_colors}, $R$ = 2.17 R$_{\odot}$ and $A_V = 0.062$ from Table \ref{table:abs_par}, and $T_{eff} = 13000$ K and $BC = -0.89$, we find the photometric parallax of this component as $\Pi = 5.747$ mas (i.e. its distance $d = 174$ pc).}

One way to assess the accuracy of the absolute parameters in Table \ref{table:abs_par}  is to compare them with those in the Gaia DR3 catalog.   An absolute magnitude of $M_{G}=0.225$ mag for the system is calculated from the distance modulus, using the $G$-band apparent magnitude $G=6.525$ mag, distance $d=172$ pc and extinction $A_{G}=0.122$ mag from the Gaia Archive\footnote{https://gea.esac.esa.int/archive/}. This absolute Gaia magnitude, $M_{G}=0.225$ mag, is then converted to the bolometric magnitude $M_{bol}=-0.595$ mag from the formula $M_{bol}$ = $M_{G}$ + $BC_{G}$, using the bolometric correction $BC_{G}=-0.820$ mag for T=13300 K, derived from  \citet{2023MNRAS.523.2440E} (their Eqn.\  4).  

{The bolometric magnitude and distance values derived from the BVR \& TESS-band LCs + HARPS RVs solutions for NO Pup A ($-0.617 \pm0.426$ mag and $171 \pm20$ pc) agree with those computed from the Gaia Archive ($-0.595\pm0.026$ mag and $172\pm1$ pc) within the error limits. }
Here, the difficulties in calibrating bolometric corrections, especially for hot stars with $T>12000$ K, should be kept in mind. 
 
\begin{table}[!tb]
\centering
\caption{Absolute parameters of the eclipsing binary NO Pup A.}
\label{table:abs_par}
\begin{tabular}{lc}
\hline
Parameter		                    & Value	              \\
\hline
$a$~(R$_{\odot}$)	                & $8.51\pm0.05$	      \\		
$e$                                 & $0.127\pm0.005$	      \\		
$i$ (deg)                           & $81.33\pm0.20$	      \\
$M_1$~(M$_{\odot}$) 	            & $3.58 \pm0.11$	  \\		
$M_2$~(M$_{\odot}$) 	            & $1.68 \pm0.09$	  \\		
$R_1$~(R$_{\odot}$) 	            & $2.17 \pm0.03$	  \\		
$R_2$~(R$_{\odot}$)	                & $1.51 \pm0.06$	  \\		
$\log g_1$	                        & $4.32 \pm0.01$	  \\		
$\log g_2$ 	                        & $4.31 \pm0.01$	  \\		
$T_1$ (K) 	                        & $13300 \pm500$	  \\		
$T_2$ (K) 	                        & $7400 \pm500$	      \\		
$L_1$~(L$_{\odot}$)                 & $133 \pm24$	  \\		
$L_2$~(L$_{\odot}$)            	    & $6.16 \pm2.16$	      \\
$M_{\rm bol,1}$ (mag) 	            & $-0.568 \pm0.193$	  \\		
$M_{\rm bol,2}$ (mag) 	            & $2.766 \pm0.380$	  \\		
$M_{\rm bol, system}$ (mag) 	    & $-0.617 \pm0.426$       \\	
$BC_{\rm V,1}$ (mag) 	            & $-0.943 \pm0.010$	  \\		
$BC_{\rm V,2}$ (mag) 	            & $0.035 \pm0.010$	  \\
$V$ (mag) 	                        & $6.490 \pm0.010$   \\
$A_V$ (mag)                         & $0.062 \pm0.031$             \\
$M_{\rm V, system}$ (mag) 	        & $0.258 \pm0.409$       \\	
$d$ (pc) from V band results        & $171 \pm20$       \\	
$BC_{\rm TESS,1}$ (mag) 	        & $-0.979 \pm0.010$	  \\		
$BC_{\rm TESS,2}$ (mag) 	        & $0.213 \pm0.010$	  \\
$TESS$ (mag) 	                    & $6.600 \pm0.010$   \\
$A_{TESS}$ (mag)                    & $0.041 \pm0.019$     \\
$M_{\rm TESS, system}$ (mag) 	    & $0.270 \pm0.413$       \\	
$d$ (pc) from TESS band results	    & $181 \pm25$       \\	
\hline
\end{tabular}
\end{table}

{We used the Geneva evolution models \citep{Yusof_etal_2022} to study the evolutionary status of the eclipsing binary NO Pup A. According to the mass values of the components listed in Table~\ref{table:abs_par}, the Geneva evolutionary tracks and isochrone were created using the interpolation interface on the Geneva group website\footnote{https://www.unige.ch/sciences/astro/evolution/en/database/}. }

{The H-R diagram shown in Figure~\ref{fig:evolution} was used to estimate the ages of the components. In this diagram, the positions of the two components are plotted on the evolutionary tracks for their measured masses. As a result, the isochrone of log(age) = 7.30 (that is, 20 My) with supersolar metallicity ($Z$=0.020) matches the positions of both components within the error limits. We may note that the atmosphere models in Section \ref{subsec:atmospheres} produced metallicities close to solar, while according to the evolution models at 20 Myr, the metallicities of the components would be somewhat larger than that.} 
{This result (age and metallicity) was also confirmed using the Padova evolution models \citep[e.g.,][]{Nguyen_2022}. }

\begin{figure}[!tb]
\centering
	\includegraphics[scale=0.55]{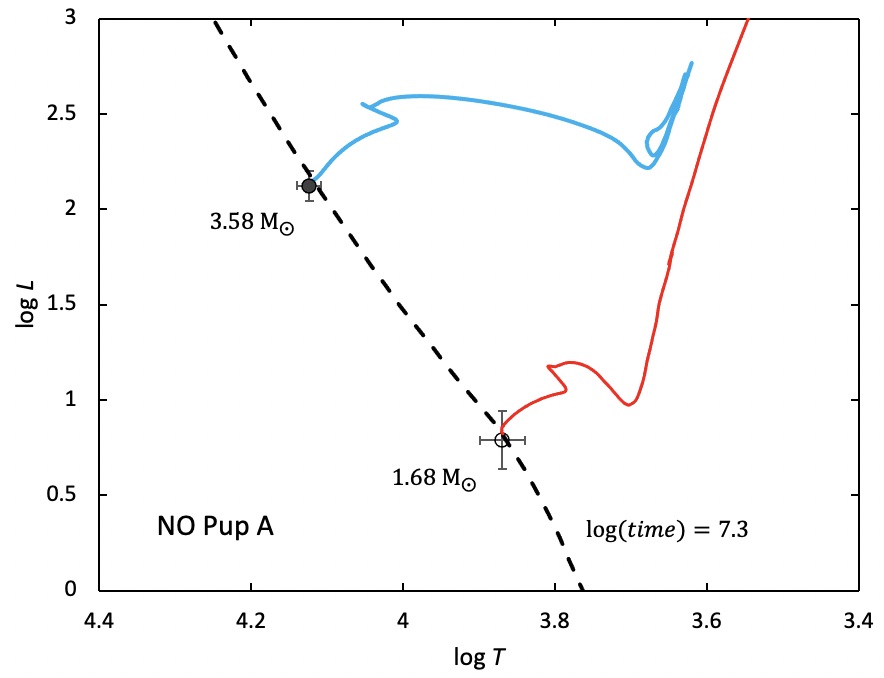}
\caption{Location of the components of NO Pup A in the H-R diagram. The Geneva evolutionary tracks for 3.58 M$_{\odot}$ (blue line) and 1.68 M$_{\odot}$ (red line), corresponding to the primary and secondary stars, are plotted for $Z_{\odot}=0.020$. The Geneva isochrone of 20 Myr for $Z$=0.020 is also indicated by the dashed black curve. Filled and open circle symbols represent primary and secondary components, respectively. Vertical and horizontal lines show error bars of the measured quantities.
}
\label{fig:evolution}
\end{figure}

\section{Astrometry}
\label{sec:astrometry}

\begin{table}
    \caption{Best-fit estimates for the orbital parameters of NO Pup B. WDS refers to the orbit parameters listed in the Washington Double Star catalogue and MCMC refers to the best-fitting parameters estimate derived by this study using Hamiltonian Markov Chain Monte Carlo. {Angles are in degrees, semi-major axis $a$ in arc-seconds, period in years, and the epoch (time of periastron passage) is in fractional Besselian year.}
    \label{tab:no_pup_b_orbit}}
    \centering
    \begin{tabular}{l|l|c}
    \hline
    Parameter                   & WDS                   & MCMC                   \\
    \hline 
     $P$ (yr)                       & 103.33            & $ 101.3 \pm 3.8 $     \\              
     $a$ (arcsec)                   & 0.199             & $ 0.179 \pm 0.012 $   \\               
     $e$                            & 0.013             & $ 0.054 \pm 0.041 $   \\      
     $\omega$ (deg)                 & 50.6              & $ 47.45 \pm 70.65 $   \\   
     $i$ (deg)                      & 136.4             & $ 154.7 \pm 13.3 $    \\
     $\Omega$  (deg)                & 277.2             & $ 318.1 \pm 73.6 $    \\   
     Epoch (yr)                     & 1990.41           & $ 1976.23 \pm 13.06$  \\
     \hline
    \end{tabular}
\end{table}

Astrometric data for the visual binary WDS J08263-39044 (NO Pup Bab) were taken from the Washington Double Star catalog (WDS, \citeauthor{Mason_2024}, \citeyear{Mason_2024}). We made use of the Markov Chain Monte Carlo (MCMC) optimisation technique as described by \citet{Ersteniuk_2024} to fit the apparent orbit. The best fit parameter estimates together with formal uncertainties are given in Table~\ref{tab:no_pup_b_orbit}, while Figure~\ref{fig:NO_Pup_B_astrometry} shows the model orbit based on these parameter values plotted against the fitted data. 

Comparison between the parameters published in the WDS from \citet{Josties_2019} and those derived from this study are in reasonable agreement, particularly for the period ($P$ in years), semi-major axis ($a$ in arcsecs), and eccentricity ($e$).  $\omega$ is the argument of periastron. $\Omega$ gives the position angle of the ascending node relative to the north direction in the sky plane. The Eulerian angles ($\omega$, inclination $i$, and $\Omega$) are less well-constrained by the MCMC fit.  

{The WDS catalog lists the primary component of NO Pup B to be about 1.42 V mag fainter than that of the B8 primary  of NO Pup A. The secondary component of NO Pup B is 0.20 V mag fainter than the primary. Using the data in Table~9.2 of \citet{Budding_2022} we can then surmise that component B likely comprises two Main Sequence dwarfs of spectral types A5 and A6 respectively. Their masses, from that same Table 9.2, are around 2 and 1.8  M$_{\odot}$. If we correspondingly adopt the mass sum of NO Pup B to be 3.8 M$_{\odot}$  we require the inclination to be close to 115 deg to allow  Kepler's third law to be satisfied with the orbit's semi-major axis being near to 34.4 AU.  Although this inclination is appreciably different from the MCMC estimate in Table~\ref{tab:no_pup_b_orbit}, the difference is comparable to that between the WDS and MCMC results.}
Further monitoring of the pair of stars in NO Pup B is clearly needed for derivation of more reliable orbital parameters (see also the related discussion on model fitting and the use of historic data in \cite{Tokovinin_2024}). 

\begin{figure}[!t]
\centering
\includegraphics[width=0.9\columnwidth]{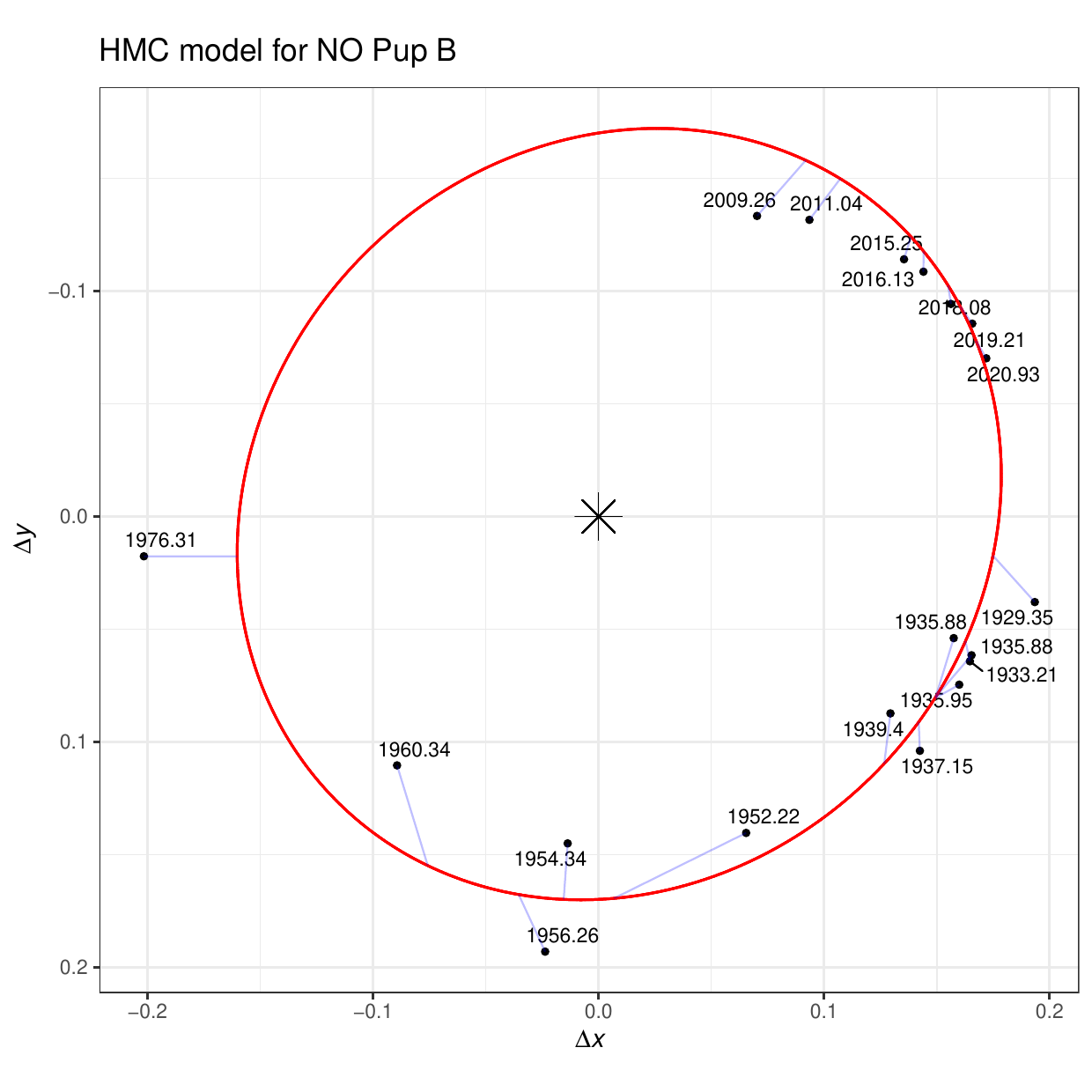}
\caption{Model fit to WDS astrometric data for NO Pup B. The red curve plots the model orbit (see Table~\protect\ref{tab:no_pup_b_orbit} for the listed parameter values), the black dots show the observational data, and the short blue lines connect the fitted data points with their expected positions along the model orbit. The primary is indicated by the star symbol at the origin. East is to the right, and down is northwards.  Labels give the observation dates.
\label{fig:NO_Pup_B_astrometry}}
\end{figure}


\section{Pulsation analysis}
\label{sec:pulsations}

 The out-of-eclipse ranges of the TESS LCs of NO Pup show systematic variations that we attributed to pulsations. We investigated these using the 120-second cadence data. With close binary variations having a dominant effect, we first separated the binary LC from the full flux. This was achieved by fitting the harmonics of the orbital frequency of NO Pup A using a method similar to the study of \citet{2023MNRAS.524..619K}. 

The residual light curve was then analysed using the \textsc{Period04} program, which derives pulsational frequencies based on a discrete Fourier transform algorithm. To estimate the significant frequencies, a 4.5$\sigma$ significance limit was applied, as outlined in the study of \citet{2021AcA....71..113B}. As a result, we obtained pulsation frequencies between 1.26 and 36.95\,d$^{-1}$. The list of derived frequencies is provided in Table\,\ref{tab:table_10}, and the amplitude spectrum is shown in Figure\,\ref{fig:power_sec}. 

    {\scriptsize
\begin{table}[!tb]
\caption{Results of the frequency analysis.}
\label{tab:table_10}
\begin{tabular}{lcccc}
\hline
                                    & Frequency       & Amp               & SNR             \\
                                    & (d$^{-1}$)      & (mmag)            &                    \\ 
                                    & $\pm$\,0.00004  & $\pm$\,0.02       &           \\
                                    \hline
                                    & High  & frequencies       &           \\

\hline
$f_1$                               &13.8868          & 0.22              & 54        \\
$f_2$                               &33.7115          & 0.19              & 45         \\
$f_3$                               &14.8767          & 0.15              & 37         \\
$f_4$                               &22.1121          & 0.10              & 16         \\
$f_5$                               &19.9876          & 0.07              & 13         \\
$f_6$                               &23.6175          & 0.04              & 8         \\
$f_7$                               &36.9543          & 0.04              & 9         \\
\hline
                                    & Low & frequencies                 &           \\
                                    \hline
$f_1$                               &1.4937          & 0.08              & 7       \\
$f_2$                               &1.2975          & 0.06              & 5       \\
$f_3$                               &1.2614          & 0.06              & 5      \\
\hline
\end{tabular}
\end{table}
}

\begin{figure}[!bt]
    \centering
    \includegraphics[width=1.0\columnwidth]{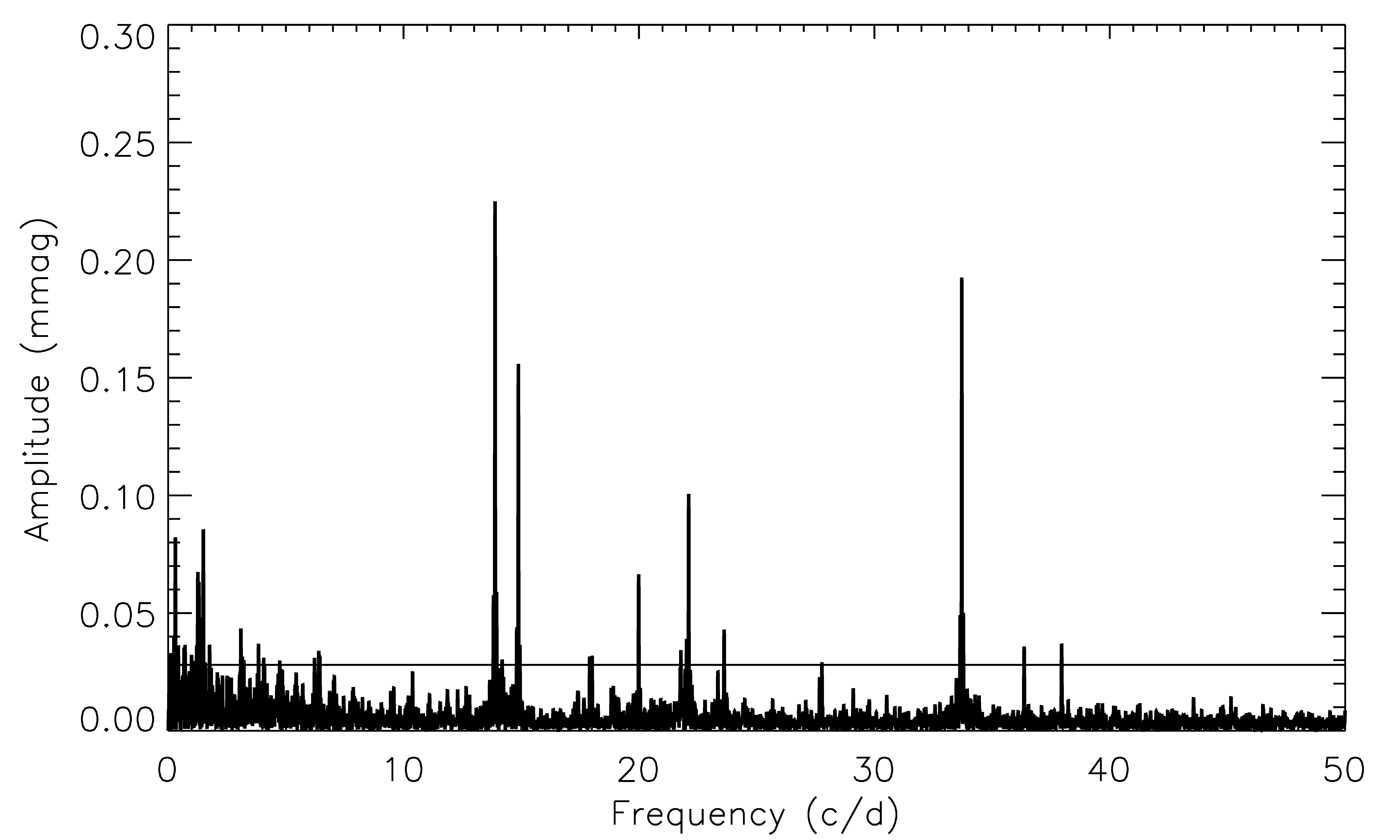}
    \caption{Power spectrum of NO\,Pup. The vertical line represents the 4.5$\sigma$ level.\label{fig:power_sec}}
\end{figure}

Taking into account the estimated frequencies and  $T_{\rm eff}$ values of the binary components, as well as the {visual companion}, we conclude that there are probably two types of oscillating star
present in the source. One is of \ds\, type, and the other exhibits pulsations characteristic of a slowly pulsating B (SPB) star. \ds\, stars, which range from A to F spectral type, show oscillations in the frequency range of approximately 5-80\,d$^{-1}$ \citep{2013AJ....145..132C}. In contrast, SPB stars are hotter objects, typically of spectral types B3-B9.  Their oscillations occur with frequencies ranging from roughly 0.3 to 1.3\,d$^{-1}$ 
(Aerts et al., 2010). These two types of pulsating stars have their own instability regions in the HR diagram, on evolving into which, oscillatory behaviour is expected from current theory.

We have plotted the close binary components Aa and Ab on the instability strips of both \ds\, and SPB stars, as shown in Figure\,\ref{fig:hr_position}. {The diagram was plotted with the same $Z$ value as Figure\,\ref{fig:evolution} assuming single-star evolution.} In Figure\,\ref{fig:hr_position}, the  component Aa lies within the SPB instability strip, while the  cooler component Ab is located within that of \ds. \, On this basis, the hotter component should be an SPB and the secondary a \ds\, star. However, we must take into account also the B component, which has mid-A spectral type components \citep{Veramendi_2014}.

\begin{figure}[t]
    \centering
    \includegraphics[width=1.0\columnwidth]{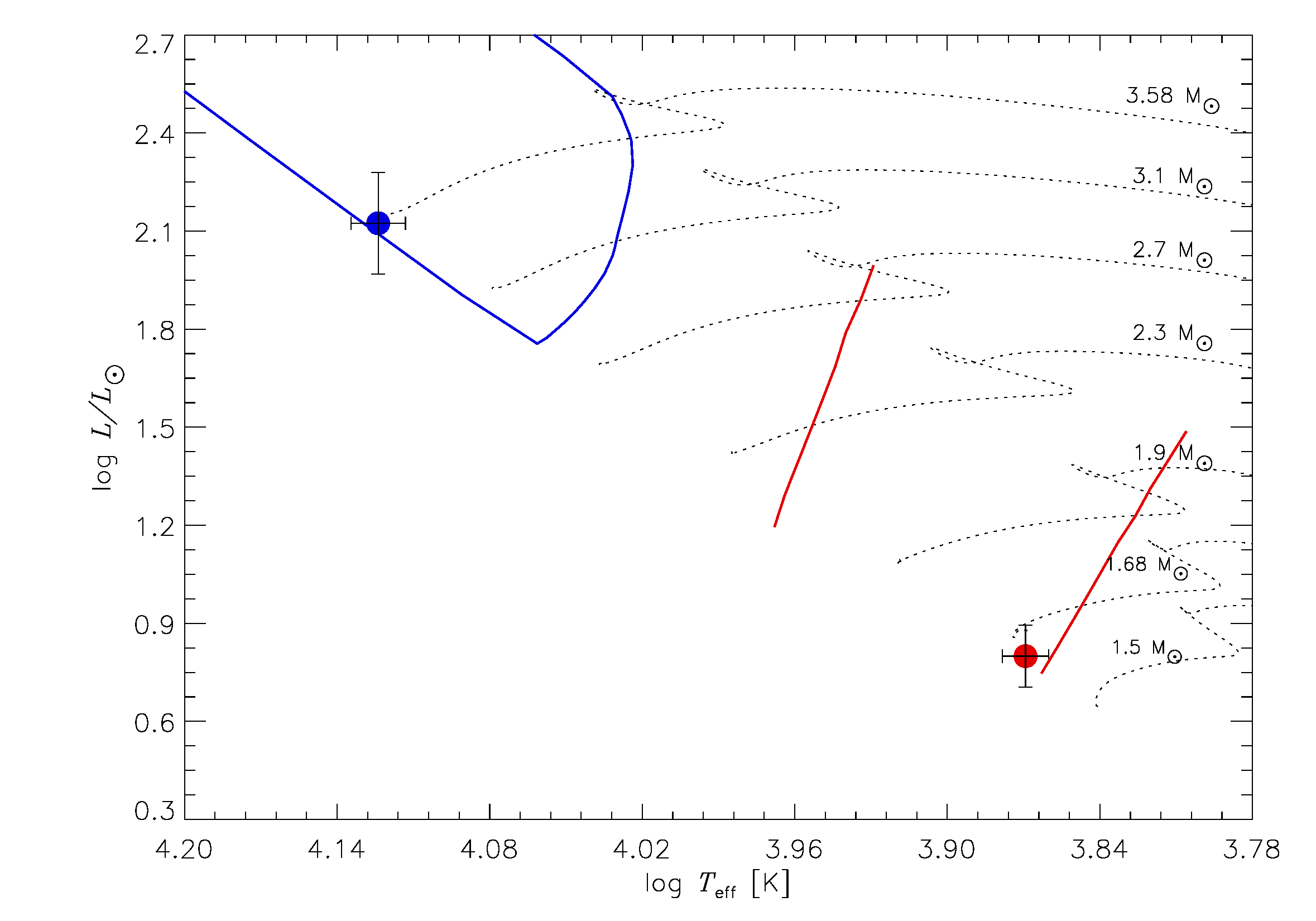}
    \caption{Position of the primary (Aa blue dot) and secondary (Ab red dot) binary components on the instability strips of \ds\, (below solid red lines \citealt{2019MNRAS.485.2380M}) and SPB (above solid blue line \citealt{1999AcA....49..119P}) stars. The theoretical evolutionary tracks (faint dotted lines) were taken from the MESA Isochrones and Stellar Tracks (MIST; \citealt{2016ApJS..222....8D}) and were generated using the same input parameters as in Figure\,\ref{fig:evolution}..\label{fig:hr_position}}
\end{figure}


Since this companion binary is located 8" away from the binary companion and the TESS pixel size is larger than this separation, we are 
at present unable to distinguish whether the \ds\, type oscillations originate from this {visual companion}. Additionally, there are hybrid \ds\,$-$\,$\gamma$\,Doradus stars \citep{2011A&A...534A.125U} that exhibit both low and high frequency pulsational effects. The component Ab could also be a \ds\,$-$\,$\gamma$\,Doradus star.  In such systems, there is generally a gap between the low-frequency $\gamma$\,Doradus-type oscillations and the higher-frequency \ds\, ones \citep{2010ApJ...713L.192G}. In our data, however, no such gap appears in the amplitude spectrum and we may exclude this possibility.

To summarise, we infer that the hotter component of the A system is likely to be an SPB object. The \ds\,-type oscillations could originate either from the cooler component of the close binary or the B component  of NO\,Pup. 

As a check,  we examined the position of the \ds\, pulsator on the proposed relationship between orbital ($P_o$) and pulsation periods ($P_{puls}$) \citep{2017MNRAS.470..915K}. This relationship was derived from data on eclipsing binaries with \ds\, pulsating components. Due to binary effects on oscillation, a correlation between pulsation and orbital periods is observed. This relationship could be used to determine whether the \ds\, pulsator is in the A or B binaries. 

The position of the \ds\, star on the log($P_o$)-log($P_{puls}$) relationship is shown in Figure\,\ref{fig:porbppuls}. As seen in the figure, the position of the \ds\, pulsator follows the general trend of the relationship. We conclude that the secondary in the eclipsing binary system, i.e.\ Ab, is probably the \ds\, variable.

\begin{figure}[t]
    \centering
    \includegraphics[width=1.0\columnwidth]{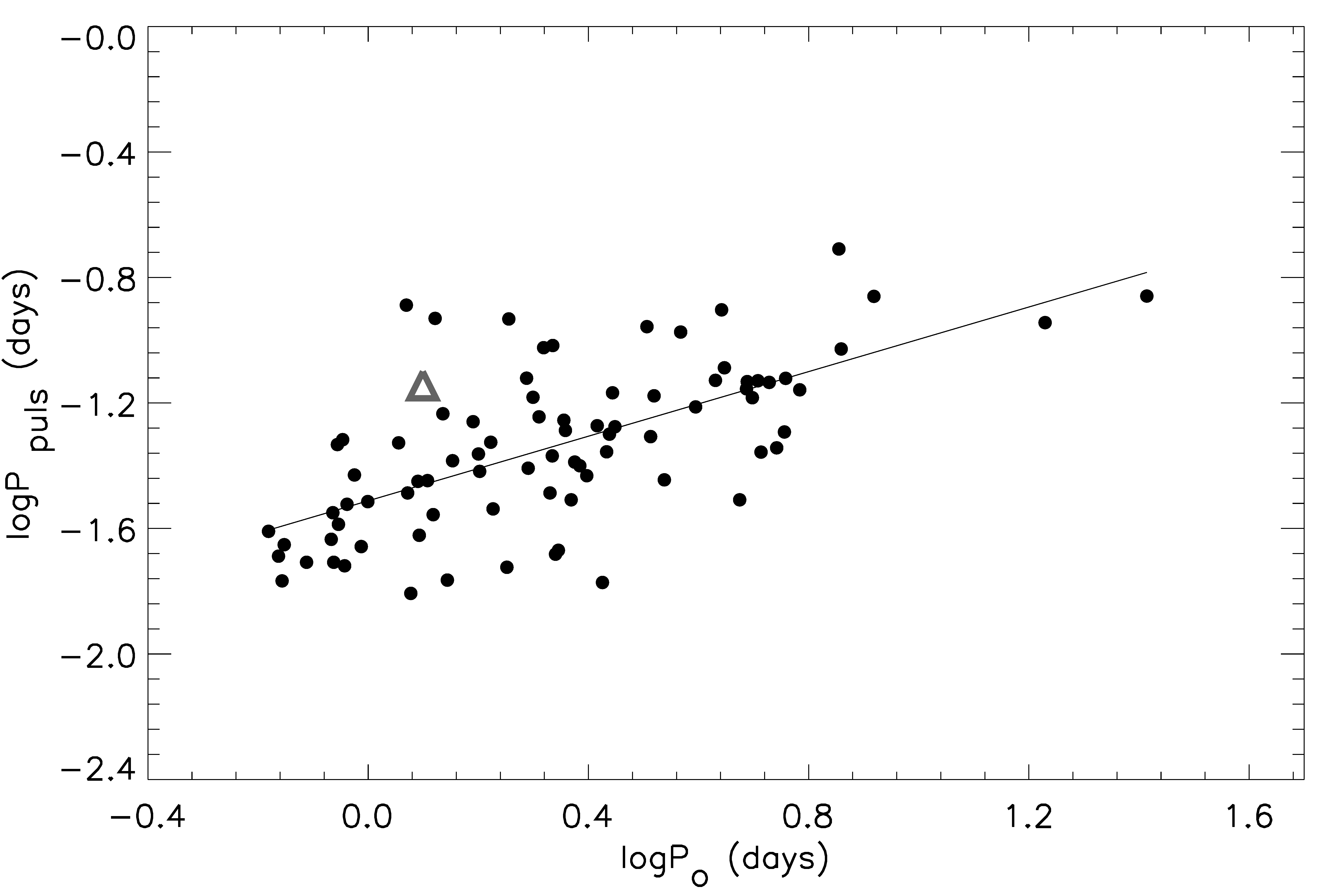}
    \caption{The log$P_o$\,-\,log$P_puls$ relationship and the position of the \ds\, star (triangle) on it. The dots represent the known \ds\, stars in eclipsing binaries, taken from \citet{2017MNRAS.470..915K}.} \label{fig:porbppuls}
\end{figure}


\section{Discussion and Concluding remarks}
\label{sec8}

NO Pup  belongs to a small, young association \citep{Tokovinin_1999}, with an estimated age of $\sim20$ Myr (see Section~\ref{sec:absolute_parameters}). The Aab component has separation {(0.03956 AU)} that is three orders of magnitude smaller than its astrometric-binary B-component. In its  present configuration, therefore, the dynamical behaviour of A is essentially governed by rotational and tidal distortions between the Aab components alone.  Given  the short period ($P \approx 1.256$ d) and close separation ($R/a \approx 0.2$), the relatively high orbital eccentricity ($e \approx 0.13$) appears anomalous. Tidal friction should tend to circularise the orbit in the available time.  On the other hand, very close encounters (of  the order of the Aab separation) with other stars in the system afford a possible way to increase orbital eccentricity ---  for example, through the  Kozai-Lidov mechanism (\cite{Naoz13}). Therefore, it is necessary to determine the timescales over which tidal friction and close 3-body encounters with Aab occur, and whether they may have a substantial effect within  the system's age.
 
The components of NO Pup B, from the information given in the WDS Catalog, are Main Sequence stars with V-magnitudes corresponding to spectral types A5V ($ M \approx 2.0M_\odot$) and A6V ($M \approx 1.8M_\odot$). From standard stellar modelling they should possess convective cores and radiative envelopes. Within these envelopes, internal ($g-$mode) gravity waves can propagate and efficiently transport angular momentum  from the convective-core boundary to the outer layers \citep{Zahn77, GN89,  Aerts10}. Such waves are partly dissipated through thermal damping in the thin, non-adiabatic surface region \citep{KK10, Townse18}. Tidal forcing \citep{Zahn75, Zahn77}) operates mostly in these outer layers with radiative damping, creating spin-up/down in the visible surface, which   may thus be a poor indicator  of the internal rotational angular momentum.
 
Orbital circularisation by tidal friction depends on the age of the system.  As a measure of its efficacy, one may specify an upper limiting value of the ratio $a/R$ of orbit size to stellar radius, below which circularisation can be achieved within an interval of one-quarter of the star's Main Sequence lifetime (\cite{Zahn77}, Table 2). For NO Pup Aab, using the 
Table~\ref{table:abs_par} masses, one finds this upper limiting value $(a/R)_{lim}$ to be $\sim 4.0$. Since for the Aa component of NO Pup, $(a/R) \sim 4.2$, the case for effective tidal circularisation of the orbit appears marginal, especially given our lack of knowledge of the rotational structure of the stars.
Zahn's analysis indicates that the upper limiting value of $a/R$ for synchronisation would be $\sim 5.6$, suggesting that tidal friction may have de-spun the components into synchronicity. The spectroscopic measurements set out in Tables \ref{tab:rv_js_fit}, \ref{tab:rv_fit} and \ref{tab:atmospar} in Section \ref{section:spectrometry}, including values of $v\sin i$, enables assessment of synchronicity of the Aab stars to the available accuracy of measurement.
{If the pseudo-synchronisation formula proposed by \citet{Hut_1981} (in his Equation 42) is used for the components of NO Pup A,  rotation velocities  95 and 66~km~s$^{-1}$ for the components Aa and Ab, respectively, are obtained. Comparison of these estimated velocity values with the measured rotation velocities (82 and 64 ~km~s$^{-1}$ for the components Aa and Ab; see Sections \ref{subsection:ucmjo_spectra} and \ref{sect:atm}), would suggest that, in this eccentric binary, the rotation and orbital motion are pseudo-synchronised for Ab but not for Aa, as noted by \citet{Veramendi_2014}. However, Aa appears to have a rotation period of 1.32 d -- close to the orbital period, as noted in Section \ref{subsection:ucmjo_spectra} -- and thus nearly synchronised to the mean orbital motion. Confirmation of these suggestions would be possible with the availability of more extensive and more accurate data.}
 
As noted in Section \ref{sec:pulsations}, the Ab component lies in the instability region of the Hertzsprung-Russell diagram for the $\delta$ Scuti variables (\cite{Aerts10}, Figure 1.12). These are likely to experience $p-$mode oscillations, whereas in the higher-mass Aa component, one would expect $g-$modes excited at the convective core/radiative envelope boundary. In addition, the high eccentricity of the orbit means that a large number of harmonics of the orbital frequency could be excited, thereby providing  possibilities for resonance with one or more of the $p-$ and $g-$ stellar oscillation modes, that in turn may be split by stellar rotation. The increase in amplitude  associated with these resonances could provide additional tidal damping. { However, harmonics of the orbital frequency $(f_{orb} = 0.7956 d^{-1})$ become separated out with the general binary  effects in the  pre-whitening process.}
  
A possible alternative explanation for the observed orbital eccentricity is by a three-body interaction. Since the orbital size of the Aab eclipsing binary system is much smaller than that of the Bab astrometric binary, we may regard Aab as a single point mass. The resulting hierarchical 3-body system may be subject to the classical Kozai-Lidov mechanism, mentioned above. Consider the Aab-Bab  system, in which  Aab is approximated by a single point mass 
The Aab component  {(mass = $M_{Aab} = 5.26 M_\odot$)} is in a wide orbit about the centre-of-mass of the Bab (astrometric) system  (masses $M_{Ba}, M_{Bb}$). The Kozai-Lidov cycle time  $t_{KL}$ for such a system, to the quadrupole level of approximation (\cite{Naoz16}, Eq (27)), is:
\begin{equation*}
    t_{KL} \approx 
    \frac{16}{30\pi}\frac{M_{tot}}{M_{Aab}}\frac{P_{wide}^2}{P_{close}}(1-e_{wide}^2)^{3/2}
\end {equation*}
where ``close'' and ``wide'' refer to the (Ba,Bb) astrometric binary and the wide orbit of Aab about Bab's centre-of-mass, respectively. $M_{tot} = M_{Ba}+M_{Bb}+M_{Aab}$ is the total mass. 

Neglecting the value of $e^2_{wide}$, and from Table \ref{table:abs_par} using  $M_{Aab} = 3.58+1.68 = 5.26 M_\odot$, then with the values given above
for the masses $M_{Ba} = 2 M_\odot, M_{Bb} = 1.8 M_\odot$, so $M_{Ba}+M_{Bb} = 3.8M_\odot$;  the distance $d$ being {171} pc, we find {$a_{close} = d\times 0.179$ arcsec $= 30.609$} AU. Now  the B-component has orbital period  {$P_{close} = 101.3$} yr, while for the wide orbit, assuming the apparent (projected) angular separation of 8.1 arcsec corresponds to the semi-major axis, we derive {$a_{wide} = 1385$} AU. Kepler's Third Law then yields {$P_{wide} \sim 17,100$} yr, on using $M_{tot} = 9.06 M_\odot$.  For the NO Pup system, we find {$t_{KL} \approx 8.5\times 10^5$} yr -- or at least an order of magnitude smaller than the estimated age of the system.\footnote{Note: Companion D (see Introduction) appears to be a highly-reddened background star, possibly belonging to the association. We assume that it is sufficiently distant not to affect the dynamics of the NO Pup system \citep{Tokovinin_1999}.} This would clearly imply the possibility of the Kozai-Lidow effect operating in the NO Pup system, to
produce apparently anomalous parameters.

\begin{figure}[!t]
    \centering
    \includegraphics[height=2.5in]{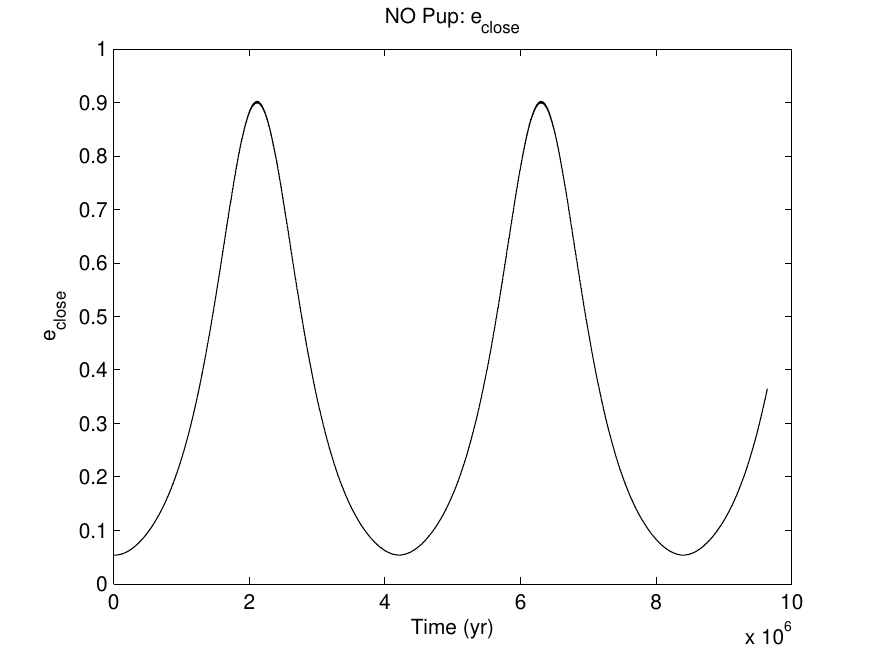} 
    \hspace{0.7cm}
    \includegraphics[height=2.5in]{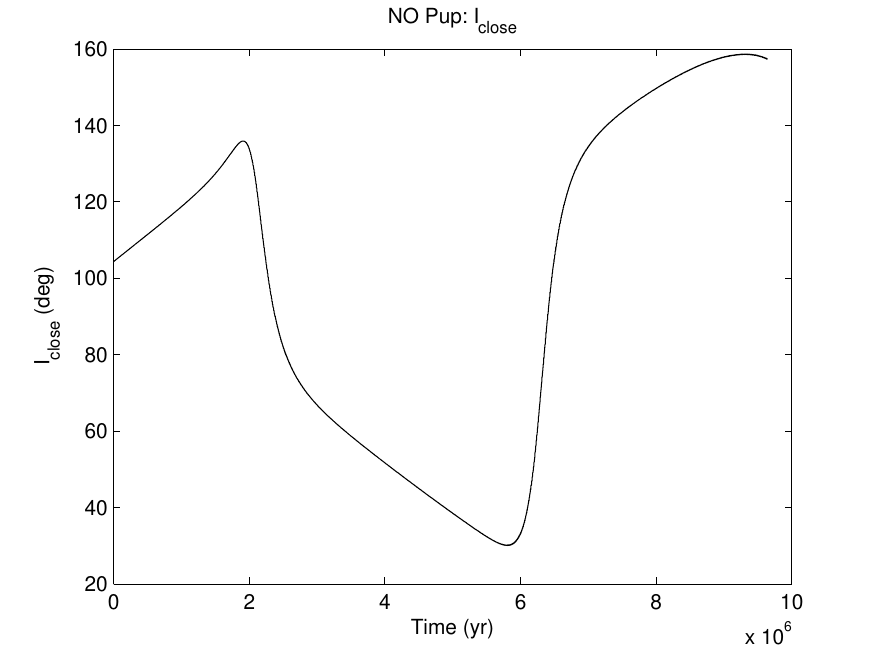}
    \caption{Top: Eccentricity of astrometric binary system (Ba,Bb) over a time 
    interval of 10 Myr, from integrating the reduced 3-body system (Aab,Ba,Bb) 
    using the Aarseth-Zare regularisation scheme. {MCMC data from Table \ref{tab:no_pup_b_orbit} was used.}The Kozai cycle is clearly 
    evident. Bottom: Inclination between the close orbital and invariable planes.}
    \label{Fig21}
\end{figure}

{In addition to classical Kozai-Lidow processes in this quadruple system's  history, since NO Pup is a member of an association \citep{Tokovinin_1999}, it may previously have comprised several more components that have since escaped via close  binary-binary or 3-body interactions (\cite{AM00, Mikkola83}). Of interest is the short-term dynamical behaviour of the system.}

{To gain insights, we may replace Aab by a point mass, as before and use the available  observational data on NO Pup. 
Initial conditions  for a 3-body model system were then chosen to be {$a_{wide} = 1385$ AU}; the orbit was assumed to be circular: $e_{wide} = 0$, and edge-on ($i_{wide} = 90 \deg$); and masses  $(M_{Ba},M_{Bb},M_{Aab}) =  (2,1.8,5.26)M_\odot$. This system was integrated over a time interval $T =  1\times 10^7$ yr, implementing an Aarseth-Zare (AZ) regularised scheme  \citep{AZ74} with   Kustaanheimo-Stiefel regularisation \citep{KS65} to  remove singularities in the equations of motion due to possible 2-body  collisions. The AZ scheme was adapted so that, after each set of 2000  time-steps of integration, the Keplerian orbital parameters corresponding to  the current smallest two of the three interparticle separations were computed.  The time transformation  $t(s)$ between ``regularised time'' $s$ and physical time $t$  was chosen to be the Lagrangian,  which gives the most accurate integration scheme and has the  advantage that the function $t(s)$ is  known explicitly \citep{Alex86}. Constancy of total energy and angular momentum was maintained  to one part in  $10^{14}$ and $10^{12}$, respectively. In addition, integrations of the Kozai-Lidov  equations, taking into account quadrupole and octupole level perturbations \citep{Naoz16} were performed. The results of AZ integrations for $e_{close}, I_{close}$ are shown in Figure~\ref{Fig21}.}
  
{Numerical integrations of the AZ scheme over 10 Myr revealed the existence 
of Kozai cycling with respect to the B-components, with a cycle time of {4.5} Myr.
 During the span of the integrations, for the astrometric binary (B-components),
 $e_{close}$ twice reached a maximum of { $e_{close} = 0.902$}, 
corresponding to a minimum separation  between the B-components 
of {3.0 AU} ($\approx 645 R_\odot$). The perturbation approach 
(quadrupole or octupole) {may become} suspect at such close encounters. However, the 
distances between Aab and both of the B components remain large ($\approx 1400$ au)
 throughout the Kozai cycle, so that tidal effects between Aab and Bab are
 negligible. By the same token, apsidal motion in the Aab eclipsing system
 will be governed exclusively by tidal and rotational distortions of its components.\\}
 
{With regard to the B-components, the quadrupole formula 
above for $t_{KL}$ yields a value $\sim 5$ times smaller than was found with the 
much more accurate AZ numerical integrations. In the immediate vicinity of 
closest approach, where $e_{close}, I_{close}$ were rapidly changing, there 
was poor agreement between the perturbative and AZ results; however, away from 
these events, there was fairly good agreement.\\}

{In this paper, we have marshalled evidence from the available data streams – photometric, spectrometric and astrometric – so as to quantify parameters as accurately as possible. This allows well-defined and comprehensive analyses. To this end we have combined various curve-fitting techniques to determine optimal values and uncertainties.  Corresponding results are presented in the preceding tables and diagrams. In turn, this permits more detailed theoretical discussion.}


\begin{acknowledgement}
This paper includes data collected by the TESS mission and obtained from the MAST data archive at the Space Telescope Science Institute (STScI). STScI is operated by the Association of Universities for Research in Astronomy, Inc., under NASA contract NAS 5–26555. This research has made use of the Washington Double Star (WDS) Catalog maintained at the U.S. Naval Observatory. We thank Dr.\ Rachel Matson for extracting data from the WDS for us, and also the University of Queensland for collaboration software. FKA thanks the  Scientific and  Technological  Research  Council (TUBITAK) project 120F330 for supporting the study. The authors thank Prof. G. Handler for his valuable comments on pulsational analysis. This work uses the VizieR catalogue access tool, CDS, Strasbourg, France; the SIMBAD database, operated at CDS, Strasbourg, France. We thank the UCMJO  time allocation committee for observing time with {\sc hercules}.  {We recently learned of the death in December 2024 of Sverre Aarseth. He was a pioneer in N-body (both small- and large N) numerical integrations and inspired many people in this field, including one of the authors  (MA). He will be greatly missed.}
\end{acknowledgement}


\paragraph{Funding Statement}

This research was partially supported by a grant from the  Scientific and Technological Research Council of T\"{u}rkiye (TUBITAK project 120F330).


\paragraph{Competing Interests}

None.


\paragraph{Data Availability Statement}

The TESS data are available from the MAST data archive (https://archive.stsci.edu/). The astrometric data may be obtained from United States Naval Observatory on request.  The other data supporting this study are available upon reasonable request to the corresponding author. 

\printendnotes

\printbibliography
\appendix 

\setcounter{table}{0}
\renewcommand{\thetable}{A\arabic{table}}

\begin{table}[t]
\caption{Radial velocity values of NO Pup derived from the HARPS spectra.} 
    \centering
    \begin{tabular}{lllllllll}
    \hline
HJD          & Orbital & RV1           & err         & RV2          & err  \\ 
2450000+     & phase   & km s$^{-1}$   & km s$^{-1}$ & km s$^{-1}$  & km s$^{-1}$ \\
\hline
426.5324  & 0.0774 & $-48.44$ & 2.77 & 178.39 & 1.85 \\ 
4888.7444 & 0.3057 & $-65.57$ & 1.34 & 225.022 & 8.88  \\ 
4890.7276 & 0.8836 & 97.42  & 3.81 & $-118.231$ & 7.45  \\
4924.4813 & 0.7387 & 146.04 & 3.78 & $-225.21$ & 6.12 \\ 
4924.5334 & 0.7802 & 142.75 & 3.06 & $-222.914$ & 2.42 \\
4924.6177 & 0.8472 & 110.46 & 1.76 & $-178.46$ & 6.83  \\ 
4924.6734 & 0.8915 & 90.77  & 5.12 & $-110.77$ & 5.78 \\ 
4924.6761 & 0.8937 & 92.77  & 2.37 & $-100.05$ & 2.15 \\ 
4924.7039 & 0.9158 & 78.01  & 3.83 & $-60.201$ & 8.13 \\ 
4925.4724 & 0.5272 & 30.64  & 1.49 &         &       \\ 
4925.4751 & 0.5294 & 32.88  & 1.75 &         &       \\
4925.5957 & 0.6253 & 110.98 & 1.25 & $-151.752$ & 1.19 \\ 
4925.5984 & 0.6275 & 110.08 & 1.85 & $-152.68$ & 3.93 \\ 
4925.6886 & 0.6993 & 139.68 & 2.25 & $-212.814$ & 2.10 \\ 
4925.6913 & 0.7346 & 130.76 & 1.47 & $-208.12$  & 5.41 \\
4925.6941 & 0.7036 & 145.60 & 4.09 & $-216.27$ & 3.53  \\ 
4925.6968 & 0.7058 & 143.13 & 5.24 & $-216.17$ & 5.87  \\ 
4926.4762 & 0.3259 & $-61.21$ & 1.92 & 212.901 & 3.76  \\ 
4926.4789 & 0.3280 & $-61.68$ & 6.85 & 216.29  & 7.40 \\
4926.5297 & 0.3685 & $-52.02$ & 4.78 & 179.24  & 5.63  \\ 
4926.5670 & 0.3981 & $-35.77$ & 1.62 & 154.56  & 6.98 \\
4926.5697 & 0.4003 & $-39.24$ & 1.81 & 148.377 & 3.15  \\ 
4926.5888 & 0.4155 & $-32.43$ & 3.42 & 133.93  & 3.27  \\
4926.5912 & 0.4174 & $-31.54$ & 2.20 & 133.763 & 3.02 \\ 
4927.4694 & 0.1161 & $-52.58$ & 1.29 & 171.91 & 2.99  \\
4927.4721 & 0.1182 & $-51.79$ & 1.56 & 174.13  & 4.75  \\ 
4927.5101 & 0.1485 & $-57.56$ & 1.88 & 194.21 & 3.06  \\ 
4927.5101 & 0.1485 & $-59.94$ & 1.28 & 193.61 & 2.63  \\
4927.5128 & 0.1506 & $-63.01$ & 1.54 & 192.18 & 5.16\\ 
4927.5503 & 0.1805 & $-64.58$ & 3.76 & 215.647 & 6.65 \\
4927.5530 & 0.1826 & $-61.76$ & 5.27 & 214.02  & 7.65  \\
4927.6230 & 0.2383 & $-71.62$ & 4.78 & 231.74  & 5.88  \\ 
4927.6257 & 0.2405 & $-70.99$ & 3.85 & 236.39 & 3.90 \\
4927.6851 & 0.2877 & $-67.16$ & 3.14 & 227.19  & 3.17 \\ 
4927.6878 & 0.2899 & $-67.36$ & 5.11 & 232.28  & 4.64  \\ 
4929.4680 & 0.7069 & 143.96 & 4.98 & $-217.33$ & 4.65 \\ 
4929.4715 & 0.7090 & 142.47 & 4.76 & $-216.042$ & 4.66 \\ 
5120.8646 & 0.9853 & 12.56  & 2.78 &         &        \\ 
5469.8902 & 0.6773 & 129.03 & 3.25 & $-194.30$ & 5.62 \\ 
5471.8856 & 0.2649 & $-66.53$ & 4.76 & 234.54  & 3.04 \\ 
6213.8976 & 0.6250 & 105.50 & 1.74 & $-119.43$ & 2.88 \\
6606.7105 & 0.1551 & $-62.68$ & 2.45 & 192.18  & 3.43 \\ 
7195.4915 & 0.6015 & 84.82  & 2.70 & $-103.13$ & 7.00 \\
\hline
    \end{tabular}
    \label{tab:rv_harps}
\end{table}

\begin{table}[!t]
\begin{center}
\caption{Identified spectral lines for NO Pup based on comparison with the ILLSS Catalogue \citep{Coluzzi_1993, Coluzzi_1999}. The lines  are confidently detected mainly for the primary (Aa). 
\label{tab:spectral_features}}
\footnotesize
\begin{tabular}{lclll}
  & & &  \\
  \hline 
\multicolumn{1}{l}{Species}  & \multicolumn{1}{c}{Order no.} &
\multicolumn{1}{l}{Adopted $\lambda$} & \multicolumn{1}{l}{Comment}  \\
\hline
He I                            & 85            &  6678.149         &   measurable, Aa                  \\
H$_{\alpha}$                    & 87            &  6562.817         &   strong, blended, Aa \& Ab       \\
Ca I                            & 88            &  6455.600         &   visible, telluric intrusions    \\
Si II                           & 89            & 6371.159          &   Aa strong, Ab weak              \\
Si II                           & 90            & 6347.091          &   Aa strong, Ab weak              \\
Fe I                            & 92            & 6157.41, 6157.734 & broad \& blended                  \\  
Si II                           & 95            & 5978.970          &  telluric intrusions              \\  
He I                            & 97            &  5875.852         &  He I triplet (av.\ $\lambda$) Aa \\
Fe II                           & 104           & 5466.020          &  weak \& blended                  \\
Fe I                            & 107           & 5315, 5321, 5329  &  visible \& blended               \\
Ca I, Fe II                     & 108           &  5264, 5275       & visible \& blended                \\
Fe II                           & 110           &  5169, 5177       &  visible \& blended               \\
Si II                           & 112, 113      & 5056.02           &    strong, Aa                     \\
He I                            & 115, 116      &  4922             & detectable,  Aa                   \\
H$_{\beta}$                     & 117           &  4861.3           &  strong, blended, Aa \& Ab        \\
He I                            & 121           &  4713             & weak,  Aa                         \\
Fe II, N III                    & 123           &  4629, 4634       &   weak                            \\
Fe I                            & 124           &  4587             &   weak                            \\
Fe II                           & 124           &  4583.8           &   weak                            \\
Fe II                           & 124           &  4576             &   weak                            \\
Fe II                           & 125           &  4549.5           &  weak \& blended                  \\
\hline 
\end{tabular}
\end{center}
\end{table}

\end{document}